\PassOptionsToPackage{table,dvipsnames}{xcolor}
\documentclass[preprints,article,accept,pdflatex,moreauthors,a4paper]{Definitions/mdpi} 
\firstpage{1} 
\pubvolume{1}
\issuenum{1}
\articlenumber{0}
\pubyear{2022}
\copyrightyear{2022}
\datereceived{} 
\dateaccepted{} 
\datepublished{} 
\hreflink{https://doi.org/} 

\usepackage{bm}
\usepackage{makecell}
\usepackage{longtable}
\usepackage{physics}

\newcolumntype{C}{>{\centering\arraybackslash}X} 

\newcolumntype{T}{>{\bfseries\centering\arraybackslash}X} 

\newcommand{\PreserveBackslashNew}[1]{\let\temp=\\#1\let\\=\temp}
\newcolumntype{Y}[1]{>{\PreserveBackslashNew\centering}p{#1}}
\newcolumntype{R}[1]{>{\PreserveBackslashNew\raggedleft}p{#1}X}
\newcolumntype{L}[1]{>{\PreserveBackslashNew\raggedright}p{#1}X}

\newcommand*{\boldrow}{%
  \global\let\rowfont\bfseries
  \bfseries}
\newcommand*{\unboldrow}{%
  \global\let\rowfont\relax
  \mdseries}
\newcommand*{\rowfont}{}

\newcommand\clearrow{\global\let\rowmac\relax}
\clearrow

\newcommand{\RNumb}[1]{\uppercase\expandafter{\romannumeral #1\relax}}

\Title{FULLY-HEAVY TETRAQUARK SPECTROSCOPY IN THE RELATIVISTIC QUARK MODEL}

\Author{Rudolf N. Faustov $^{1}$, Vladimir O. Galkin $^{1}$ and Elena M. Savchenko $^{1,2,}$*}

\AuthorNames{Rudolf N. Faustov, Vladimir O. Galkin and Elena M. Savchenko}

\AuthorCitation{Faustov, R.N.; Galkin, V.O.; Savchenko, E.M.}

\address{$^{1}$ \quad Federal Research Center ``Computer Science and Control'', Russian Academy of Sciences, Vavilov Street 40, 119333 Moscow, Russia
\\
$^{2}$ \quad Faculty of Physics, M.V.Lomonosov Moscow State University, Leninskie Gory 1-2, 119991 Moscow, Russia}

\corres{Correspondence: savchenko.em16@physics.msu.ru}

\abstract{Masses of the ground and excited (1P, 2S, 1D, 2P, 3S) states of the fully-heavy tetraquarks, composed of charm ($c$) and bottom ($b$) quarks and antiquarks, are calculated in the diquark-antidiquark picture within the relativistic quark model based on the quasipotential approach and quantum chromodynamics. The quasipotentials of the quark-quark and diquark-antidiquark interactions are constructed similarly to the previous consideration of mesons and baryons. Relativistic effects are consistently taken into account. A tetraquark is considered as a bound state of a diquark and an antidiquark. The finite size of the diquark is taken into account, using the form factors of the diquark-gluon interaction. It is shown that most of the investigated states of tetraquarks lie above the decay thresholds into a meson pair, as a result they can be observed  only as broad resonances. The narrow state X(6900) recently discovered in the di-$J/\psi$ production spectrum by the LHCb, CMS and ATLAS Collaborations corresponds to an excited state of the fully-charmed tetraquark. Other recently discovered  exotic heavy resonances X(6200), X(6400), X(6600), X(7200), X(7300) can also be interpreted as the different excitations of the fully-charmed tetraquark.}


\begin{document}

\section{Introduction}
\label{SecIntr}

\par
The quark model of hadrons predicts various possible stable combinations of valence quarks and antiquarks, but for many decades only two kinds of combinations were observed: baryons, consisting of three quarks ($qqq$), and mesons, consisting of a quark and an antiquark ($q\bar q$). Other possible combinations such as tetraquarks ($qq\bar q\bar q$), pentaquarks ($qqqq\bar q$), glueballs ($gg$), hybrids ($q\bar q g$) and others were called ``exotic''.
\par
For many years the very existence of those states was unclear, since there was no convincing experimental evidence for them. The first reliable candidate for an exotic state was the X(3872) particle (Belle 2003~\cite{X-3872-2003}). This is a charmonium-like state with an extremely narrow width ($\Gamma=1.19 \pm 0.21$ Mev~\cite{PDG2022}) and uncharacteristic decays breaking the isospin ($\frac{Br(X(3872) \rightarrow \omega J/\psi)}{Br(X(3872) \rightarrow \pi^{+}\pi^{-} J/\psi)} = 1.1 \pm 0.4$~\cite{X-3872-2003, X-3872-2004}). Thus, the X(3872) does not fit into the naive quark picture of hadrons except as in the form of the
two-quark--two-antiquark state ($cu\bar c\bar u$). Soon after, the first explicitly exotic state $Z_{c}^{\pm}$(4430) (LHCb 2014~\cite{Z-4430-2014}) was discovered. This particle is of special interest since it is the charged charmonium state. A nonzero electric charge means that, in addition to a pair of charmed quark and antiquark, it contains also a  light quark and antiquark of different flavors ($cu \bar c \bar d$, $cd \bar c\bar u$). Currently  a few dozen of candidates and reliably confirmed tetraquarks ($cc \bar c \bar c$ –- X(6900) -- LHCb 2020~\cite{U10-ccLHCb2020}, CMS 2022~\cite{ccCMS2022}, ATLAS 2022~\cite{ccATLAS2022}, etc.) and pentaquarks ($uudc \bar c$ -- $P_{c}^{+}$(4380),  $P_{c}^{+}$(4450) -- LHCb 2015~\cite{Penta2015}) have been discovered.  The most recent detailed review can be found in Ref.~\cite{Review2022}.
\par
The unified theoretical picture of exotic states has not been developed yet. In the absence of a direct description of hadrons from first principles of QCD, theorists have to use model assumptions about the structure and nature of the interaction of quarks in exotic hadrons. As a result, there are theoretical approaches that assume a different composition of exotic states and methods for their nonperturbative description. The predictions obtained within their framework agree with experimental data with varying degrees of success. The object of our research from all exotic states are fully-heavy tetraquarks, consisting of two heavy quarks and two heavy antiquarks. This choice significantly reduces the number of approaches applicable for their description. At the moment, there are already a number of theoretical calculations within the framework of different models, but there is no consensus on which of the predicted states are long-living enough for their experimental detection.
\par
Experimental searches for such states are actively conducted at Large Hadron Collider (LHC) in CERN. At present, the LHCb~\cite{U10-ccLHCb2020, U24LHCBbb-bbNotLHCb2018}, CMS \cite{bbNotCMS2017, U25CMSbb-bbNotCMS2020, ccCMS2022} and ATLAS~\cite{ccATLAS2022} Collaborations are actively searching for the fully-charmed $cc\bar c\bar c$ and fully-bottom $bb\bar b\bar b$ tetraquarks. The fully-charmed states $cc\bar c\bar c$ are searched as the intermediate resonances in the processes $p + p \rightarrow J/\psi(1S) J/\psi(1S)$, $p + p \rightarrow J/\psi(1S) \psi(2S)$and $p + p \rightarrow J/\psi \mu^{+}\mu^{-}$ at $\sqrt{s} = 7, 8$ and $13$ TeV (LHCb). The predicted mass of the $cc\bar c\bar c$ tetraquark lies in the range of $5.8-7.4$ GeV. Searches for it were performed in the mass range $6.2-7.4$ GeV. In 2020 the LHCb Collaboration announced the discovery of the narrow resonance X(6900) in di-$J/\psi$ spectrum~\cite{U10-ccLHCb2020}, which, according to the measured mass and width, is a candidate for the excited $cc\bar c\bar c$ state. Also several other broad structures  peaking at about $6.4$ and $7.2$ GeV were reported. They can  be other excitations of the same $cc\bar c\bar c$ tetraquark. Later in 2022 CMS~\cite{ccCMS2022} and ATLAS~\cite{ccATLAS2022} Collaborations presented preliminary data confirming X(6900) and giving hints of a few more states including structures at $6.4$ and $7.2$ GeV. 
\par
In the sector of fully-bottom tetraquarks $bb\bar b\bar b$ there is no progress yet. These tetraquark states are searched as the intermediate resonances in the processes $p + p \rightarrow \Upsilon(1S) \Upsilon(1S)$ and $p + p \rightarrow \Upsilon \mu^{+}\mu^{-}$ at $\sqrt{s}=7, 8$ and $13$ TeV (LHCb) and 8 and 13 TeV (CMS). The predicted mass of the $bb\bar b\bar b$ state lies in the range of $18.4-18.8$ GeV. Searches for the $bb\bar b\bar b$ state were carried out in the mass range of $17.5-20.0$ GeV (LHCb) and $17.5-19.5$ GeV (CMS). CMS also searched for the narrow resonances in the mass range  $16.5-27$ GeV. However,  none of these studies revealed reliable signs of a resonance with properties  expected for the exotic $bb\bar b\bar b$ state in the given process and at such energies.
\par
The paper is organized as follows. In Sec.~\ref{SecModel} we give a description and physical justification of the model for  studying  these tetraquark structures. In Sec.~\ref{SecTheor} we describe the relativistic quark model and its application to the calculation of the tetraquark mass spectra. In Sec.~\ref{SecRes} we present the results of our calculations. In Sec.~\ref{SecThr} we analyze our predictions comparing them with the thresholds for the strong fall-apart decays. In Sec.~\ref{SecComp} we give a comparison of our results with the predictions of other scientific groups. Finally, in  Sec.~\ref{SecCon}  the results and conclusions are summarized.

\section{Model description of fully-heavy tetraquarks}
\label{SecModel}

\par
Tetraquark is a bound state of two quarks and two antiquarks. There are 6 flavors of quarks, and according to their masses they can be divided into two groups: light (with the current masses less than the $\Lambda_{\text{QCD}} \approx 200$ MeV, quark confinement energy) and heavy (with masses larger than $\Lambda_{\text{QCD}}$) quarks. Light quarks are the $u$-quark with mass $2.16^{+0.49}_{-0.26}$ MeV, $d$-quark with mass $4.67^{+0.48}_{-0.17}$ MeV and $s$-quark with mass $93.4^{+8.6}_{-3.4}$ MeV. Heavy quarks are the $c$-quark with mass $1.27 \pm 0.02$ GeV, $b$-quark with mass $4.18^{+0.03}_{-0.02}$ GeV and $t$-quark with mass $172.69 \pm 0.30$ GeV~\cite{PDG2022}. We will focus on the fully-heavy tetraquarks. However, the t-quark is special. It is almost two orders of magnitude heavier than other heavy quarks, and thus it quickly decays via the weak interaction, not having enough time to form a bound state~\cite{Lifetime1986}. Therefore, we will not consider it.
\par
From the two flavors of quarks and antiquarks, many combinations can be made. We have already done calculations for the ground states masses for all possible compositions~\cite{Savch2020, Savch2021}. However, given the large number of possible excited states, it is more rational to select and study  those combinations that are easier to detect experimentally. The most convenient of these are the symmetric compositions: fully-charmed $cc \bar c \bar c$, doubly charmed-bottom $cb \bar c \bar b$, and fully-bottom $bb \bar b \bar b$ tetraquarks. The reason for the preference of such combinations is that the tetraquarks are formed from the closely produced quark and antiquark pairs. Thus the formation of these states requires the production of only two pairs ($2 \times c\bar c$, $c\bar c + b\bar b$ and $2 \times b\bar b$) while the formation of other combinations requires the production of at least three  pairs, which is a less probable event.
\par
We consider the tetraquark as a bound state of a diquark $QQ'$ and an antidiquark $\bar Q \bar Q'$. This model is not new and is widely used in the hadron spectroscopy, giving good agreement between the calculations (for example, baryon masses) and experiments. Also theoretically predicted spectrum of possible baryon excitations in the genuine three body picture is much wider than the experimentally observed one. The quark-diquark model of baryons, on the other hand, freezes some degrees of freedom and  imposes the necessary restrictions that bring the theory into better agreement with experiment~\cite{Baryo2005, Regge2011}.
\par
Another widely used model for the tetraquarks description is a molecular picture. We consider such a picture of  fully-heavy tetraquarks significantly less probable. Indeed, in this case the meson molecule model has the following main problems. The interaction between  mesons in a molecule is  either due to the Van der Waals forces, or through the exchange of another meson containing the same quarks as in the molecule. The Van der Waals forces are weak in general and cannot provide sufficient binding.  In  the fully-heavy tetraquarks only heavy mesons can be exchanged: $c\bar c, c\bar b, b\bar c, b\bar b$. Such interaction is described by the Yukawa potential and its strength decreases with the increasing mass of the exchanged meson. Therefore, such potential can provide a weak coupling in the case of the exchange of the light mesons, like pions ($M_{\pi^{\pm}}=139.57$ MeV~\cite{PDG2022}), but in the considered case ($M_{min}=M_{\eta_{ c}}=2983.9 \pm 0.4$ MeV~\cite{PDG2022}) the coupling will be vanishingly small.
\par
In the diquark consideration one must  take into account that a (anti)diquark is a bound system of fermions, and therefore must obey the generalized Pauli principle: the complete wave function of a (anti)diquark must be antisymmetric. The diquark color representation can be either antitriplet (the antisymmetric color wave function) or sextet (the symmetric color wave function). But in the case of sextet the interaction potential between the quarks within the diquark is repulsive and thus corresponding diquark cannot be a bound state, which we consider inappropriate for our problem. The above argument applies to the antidiquark.  In the following we consider only color antitriplet diquarks. This means that if a (anti)diquark is composed of (anti)quarks of the same flavor (the symmetric flavor wave function), it can only have the symmetric spin wave function, thus being in the axialvector (A) state. If a diquark consists of quarks of different flavors, it can be either in the axialvector (A) or scalar (S) state.

\section{Relativistic diquark-antidiquark model}
\label{SecTheor}

\par
For the calculation of the masses of tetraquarks, we use the relativistic quark model based on the quasipotential approach and the diquark-antidiquark picture of tetraquarks. In this approach the masses of tetraquarks are the solutions of the relativistic Schr\"odinger-type quasipotential equation~\cite{Logunov1963, Faustov1985, TMF1990}. This equation describes the bound state of two particles in a given quasipotential. We first apply it to the quark-quark system forming a diquark and then to the diquark-antidiquark system forming a tetraquark~\cite{U27Bach42misc, Diquark2006}:
\begin{equation}
\label{EqKv}
\bigg( \frac{b^{2}(M)}{2\mu_{R}(M)}-\frac{\mathbf{p}^{2}}{2\mu_{R}(M) } \bigg) \Psi_{T, d}(\mathbf{p}) = \int \frac{d^{3}q}{(2\pi)^{3}}\: V(\mathbf{p}, \mathbf{q}; M) \Psi_{T, d}(\mathbf{q}).
\end{equation}
Here
$\mathbf{p}$ is a vector of the relative momentum, $M$ is the mass of the bound state, $\mu_{R}$ is the relativistic reduced mass of the constituents given by
\begin{equation}
\label{EqMu}
\mu_{R} = \frac{E_{1}E_{2}}{E_{1}+E_{2}} = \frac{M^4-(m_{1}^{2} - m_{2}^{2})^{2}}{4M^{3}},
\end{equation}
where $m_{1, 2}$ are masses of the constituents and $E_{1, 2}$ are the on-mass-shell energies of constituents:
\begin{equation}
\label{EqEn}
E_{1, 2} = \frac{M^{2}-m_{2, 1}^{2}+m_{1, 2}^{2}}{2M}.
\end{equation}
$b^{2}(M)$ is the on-mass-shell relative momentum in the center-of-mass system squared:
\begin{equation}
\label{EqB^2}
b^{2}(M)  = \frac{[M^{2}-(m_{1}+m_{2})^{2}][M^{2}-(m_{1}-m_{2})^{2}] }{4M^{2}},
\end{equation}
$\Psi_{T, d}(\mathbf{p})$ are the bound state wave functions, $V(\mathbf{p}, \mathbf{q}; M)$ is the quasipotential operator  of the constituents.
\par
The equation~(\ref{EqKv}) is relativistic. On the left hand side it contains relativistic kinematics: the reduced mass of the bound state $\mu_{R}$ and the on-mass-shell relative momentum $b^{2}(M)$ are functions of the bound state mass  $M$ (Eq.~(\ref{EqMu})). The relativistic dynamics is contained on the right hand side of the Eq.~(\ref{EqKv}), in the quasipotential $V(\mathbf{p}, \mathbf{q}; M)$. The quasipotential is constructed with the help of the off-mass-shell scattering amplitude, projected onto the positive-energy states and contains all relativistic spin-independent and spin-dependent contributions.
\par
Constructing the quasipotential of the quark-quark interaction, we assume that the effective interaction is the sum of the usual one-gluon exchange term with the mixture of the long-range vector and scalar linear confining potentials, where the vector confining potential vertex contains the additional Pauli term. Due to the difference in the $QQ'$ and $\bar{Q} \bar{Q'}$ color states the quark-quark interaction quasipotential is considered to be ${V_{QQ'} = \frac{1}{2} V_{\bar{Q} \bar{Q'}}}$ of the quark-antiquark interaction quasipotential \cite{Baryo2005} and is given by 
\begin{equation}
\label{EqPot}
V(\mathbf{p}, \mathbf{q}; M) = \overline{u}_{1}(p) \overline{u}_{2}(-p) \mathcal{V}(\mathbf{p}, \mathbf{q}; M) u_{1}(q) u_{2}(-q),
\end{equation}
with
\begin{align}
\mathcal{V}(\mathbf{p}, \mathbf{q}; M) & = \mathcal{V}_{\rm OGE} + \mathcal{V}_{\rm conf.}^{V} + \mathcal{V}_{\rm conf.}^{S} \nonumber
\\
& = \frac{1}{2} \bigg[ \underbrace{\frac{4}{3}\alpha_{s}D_{\mu\nu}(\mathbf{k})\gamma_{1}^{\mu}\gamma_{2}^{\nu}}_{\substack{\rm one-gluon-exchange}} + \; \underbrace{V^{V}_{\rm conf.}(\mathbf{k})\Gamma_{1}^{\mu}(\mathbf{k})\Gamma_{2;\mu}(-\mathbf{k}) + V^{S}_{\rm conf.}(\mathbf{k})}_{\substack{\rm confinement}} \bigg]. \label{EqMatrElqq}
\end{align}
Here $\mathbf{k}=\mathbf{p}-\mathbf{q}$, $\gamma_{l}^{\mu,\nu}$ and $u_{l}^{\lambda}(p)$ are the Dirac matrices and spinors:
\begin{equation}
\label{EqDir}
u_{l}^{\lambda}(p) = \sqrt{\frac{\varepsilon_{l}(p)+m_{l}}{2\varepsilon_{l}(p)}} \begin{pmatrix} 1 \\ \frac{\bm{\sigma}\mathbf{p}}{\varepsilon_{l}(p) + m_{l}} \end{pmatrix} \chi^{\lambda}, \quad l=1,2,
\end{equation}
where $\varepsilon_{l}(p)$ is the quark energy:
\begin{equation}
\label{EqEnQ}
\varepsilon_{l}(p) = \sqrt{m_{l}^{2} + \mathbf{p}^{2}}, \quad l=1,2,
\end{equation}
$\bm{\sigma}$ and $\chi^{\lambda}$ are the Pauli matrices and spinors:
\begin{equation}
\label{EqPauli}
\chi^{\lambda} = 
\begin{pmatrix}
  1 \\ 
  0
\end{pmatrix}, 
\begin{pmatrix}
  0 \\ 
  1
\end{pmatrix}, \quad \lambda=1,2.
\end{equation}
$\alpha_{s}$ is the running QCD coupling constant with freezing~\cite{alphasSimonov, alphasBadalian}:
\begin{gather}
\alpha_{s}(\mu^{2}) = \frac{4\pi}{(11-\frac{2}{3}\eta_{f}) \: \text{ln} [\frac{\mu^{2}+ M_{BG}^{2}}{\Lambda^{2}}]}, \label{EqAlpha}
\\[3pt]
\begin{cases}
	\mu = \frac{2m_{1}m_{2}}{(m_{1}+m_{2})}, \\[3pt]
	M_{BG}=2.24\sqrt{A}=0.95 \; \rm GeV, \\
	\Lambda = 414 \; \rm MeV, \\
	\eta_{f} = 
		\begin{cases}
			4, \quad Q=Q'=c, \\
			5, \quad Q,Q'=c,b,
		\end{cases}
\end{cases} \label{EqAlphaPar}
\end{gather}
where the scale $\mu$ is chosen to be equal to the reduced constituents mass, $M_{BG}$ is the background mass, $\Lambda$ is the parameter of the running coupling constant obtained from the analysis of meson mass spectra and $\eta_{f}$ is the number of open flavors. $D_{\mu\nu}(\mathbf{k})$ is the gluon propagator in the Coulomb gauge:
\begin{equation}
\label{D}
\begin{cases}
	D^{00}(\mathbf{k}) = - \frac{4\pi}{\mathbf{k}^{2}}, \\
	D^{ij}(\mathbf{k}) = - \frac{4\pi}{k^{2}} \bigg( \delta^{ij} - \frac{k^{i}k^{j }}{\mathbf{k}^{2}} \bigg), \\
	D^{0i}=D^{i0}=0,
\end{cases} \quad i,j=\overline{1,3}.
\end{equation}
$\Gamma_{l}^{\mu}$ is the effective long-range vector interaction vertex \cite{Param1992}, it contains both Dirac and Pauli terms:
\begin{equation}
\label{EqGamma}
\Gamma_{l;\mu}(\mathbf{k}) = \gamma_{\mu}+\frac{i\kappa}{2m_{l}}\sigma_{\mu\nu} \tilde{k}^{\nu}, \quad \tilde{k}=(0, \mathbf{k}), \quad l=1,2,
\end{equation}
where $\sigma_{\mu\nu}$ is the commutator of the Dirac matrices, $\kappa$ is the long-range anomalous chromomagnetic moment of quarks and $\frac{i\kappa}{2m_{l}}\sigma_{\mu\nu}\tilde{k}^{\nu}$ is the anomalous chromomagnetic interaction.
$V_{\text{conf.}}^{\text{v, s}}$ are the vector and scalar confining potentials which in the nonrelativistic limit in configuration space (consistent with the lattice calculations) have the form
\begin{gather}
V_{\rm conf.}^{V}(r) = (1-\varepsilon)V_{\rm conf.}(r), \nonumber
\\
V_{\rm conf.}^{S}(r) = \varepsilon V_{\rm conf.}(r), \nonumber
\\
V_{\rm conf.}^{V}(r)+V_{\rm conf.}^{S}(r) = V_{\rm conf.}(r ) = Ar+B, \label{EqConf}
\end{gather}
where $\varepsilon$ is the mixing coefficient. Therefore in the nonrelativistic limit the $QQ'$ quasipotential reduces to:
\begin{equation}
\label{EqVnr}
V_{QQ'}^{\rm NR}(r)=\frac{1}{2} V_{Q\bar Q'}^{\rm NR}(r) = \frac{1}{2} \Big( -\frac{4}{3} \frac{\alpha_{s}}{r} +Ar+B \Big),
\end{equation} 
reproducing the usual Cornell potential. Thus, our quasipotential can be viewed as its relativistic generalization. It contains both spin-independent and spin-dependent relativistic contributions. 
\par
Constructing the diquark-antidiquark quasipotential, we use the same assumptions about the structure of the short- and long-range interactions. We also take into account the finite size of the diquarks and their integer spin. The quasipotential then is given by~\cite{Diquark2006, Diquark2007}:
\begin{align}
V(\mathbf{p}, \mathbf{q}; M) & = \underbrace{\frac{<d({\mathcal{P})}|J_{\mu}|d(\mathcal{Q})> }{2\sqrt{E_{d}}\sqrt{E_{d}}} \frac{4}{3}\alpha_{s}D^{\mu\nu}(\mathbf{k}) \frac {<d'({\mathcal{P'})}|J_{\nu}|d'(\mathcal{Q}')>}{2\sqrt{E_{d'}}\sqrt{E_{d '}}}}_{\substack{\rm diquark-gluon \; interation}} \nonumber
\\
& + \underbrace{\Psi_{d}^{*}(\mathcal{P})\Psi_{d'}^{*}(\mathcal{P'}) [J_{d; \mu}J_{d'}^{\mu}V_{\rm conf.}^{V}(\mathbf{k})+V_{\rm conf.}^{S}(\mathbf{k})]\Psi_{d}(\mathcal{Q})\Psi_{d'}(\mathcal{Q}')}_{\substack{\rm confinement}}. \label{EqPotdd}
\end{align}
Here $d$ and $d'$ denote the diquark and antidiquark, ${\mathcal{Q}^{(')}=(E_{d^{(')}} \; \pm\mathbf{q})}$ and ${\mathcal{P}^{(')}=(E_{d^{(')}}, \; \pm \mathbf{p})}$ are the initial and final diquark momenta respectively, $k=\mathcal{P}-\mathcal{Q}$, $E_{d, d'}$ are the on-shell diquark energies (similar to Eq.~(\ref{EqEn})):
\begin{equation}
\begin{cases}
	E_{d}=\frac{M^{2}-M_{d'}^{2}+M_{d}^{2}}{2M}, \\[5pt]
	E_{d'}=\frac{M^{2}-M_{d}^{2}+M_{d'}^{2}}{2M}, 
\end{cases} \label{EqEd}
\end{equation}
where $M_{d, d'}$ are the diquark and antidiquark masses. $\Psi_{d}(p)$ is the diquark wave function:
\begin{equation}
\label{EqPsid}
\Psi_{d}(p)=
	\begin{cases} 
		1, \quad & \rm scalar \\ 
		\epsilon_{d}(p), \quad & \rm axialvector 
	\end{cases} \quad \rm diquarks,
\end{equation}
where $\epsilon_{d}(p)$ is the polarization vector of an axialvector diquark with momentum $\mathbf{p}$:
\begin{gather}
\epsilon_{d}(p)=\Bigg( \frac{(\bm{\epsilon_{d}\mathbf{p}})}{M_{d}}, \; \bm{\epsilon_{d}} + \frac{(\bm{\epsilon_{d}\mathbf{p}})\bm{\mathbf{p}}}{M_{d}(M_{d}+E_{d}(p))} \Bigg), \label{EqEpsd}
\\
\epsilon_{d}^{\mu}(p)p_{\mu}=0, \nonumber
\end{gather}
where $E_{d}(p)$ is the diquark energy (similar to Eq.~(\ref{EqEnQ})):
\begin{equation}
\label{EqEnD}
E_{d}(p)=\sqrt{M_{d}^{2}+\mathbf{p}^{2}}.
\end{equation}
$J_{d; \mu}$ is the effective long-range vector interaction vertex of the diquark:
\begin{gather}
J_{d; \mu} = 
	\begin{cases} 
		\frac{(\mathcal{P}+Q)_{\mu}}{2\sqrt{E_{d}}\sqrt{E_{ d}}}, \quad & \rm scalar \\[5pt]
		-\frac{(\mathcal{P}+Q)_{\mu}}{2\sqrt{E_{d}}\sqrt{E_{d}}}+\frac{i \mu_{d}}{2M_{d}} \sum\nolimits_{\mu}^{\nu} \tilde{k}_{\nu}, \quad & \rm axialvector 
	\end{cases} \quad \rm diquarks, \label{EqJd}
\\
\mu_{d} = 0, \label{EqMud}
\end{gather}
where $\mu_{d}$ is the total chromomagnetic moment of the diquark, which we choose equal to zero to vanish the long-range chromomagnetic interaction. $(\sum\nolimits_{\rho\sigma})_{\mu}^{\nu}$ is a fully antisymmetric tensor:
\begin{equation}
\label{EqSigma}
(\sum\nolimits_{\rho\sigma})_{\mu}^{\nu}=-i(g_{\mu\rho}\delta_{\sigma}^{\nu}-g_{\mu\sigma}\delta_{\rho}^{\nu}).
\end{equation}
$<d(\mathcal{P})|J_{\mu}|d(\mathcal{Q})>$ is the diquark-gluon interaction vertex (Fig.~\ref{OGE}), which accounts for the internal structure of the diquark and leads to the emergence of the form factor $F(r)$ smearing the one-gluon exchange potential~\cite{U27Bach42misc}:

\begin{figure}[H]
\center{\includegraphics[width=6.5 cm]{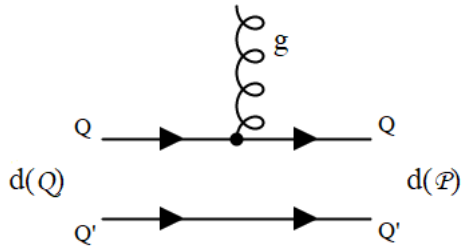}}
\caption{Feynman diagram of the gluon emission by a quark in a diquark.}
\label{OGE}
\end{figure}

\begin{equation}
\label{EqCurrent}
<d(\mathcal{P})|J_{\mu}|d(\mathcal{Q})> = \int \frac{d^{3}p \; d^{3}q}{(2\pi)^{6}} \; \overline{\Psi}_{d}^{\mathcal{P}}(\mathbf{p}) \Gamma_{\mu}(\mathbf{p}, \mathbf{q}) \Psi_{d}^{\mathcal{Q}}(\mathbf{q}).
\end{equation}
Here $J_{\mu}$ is the quark current:
\begin{equation}
\label{EqJmu}
J_{\mu} = \bar Q \gamma_{\mu} Q,
\end{equation}
where $Q, \bar Q$ denotes the initial and final states of the quark, respectively. $\Gamma_{\mu}(\mathbf{p}, \mathbf{q})$ is the vertex function of the diquark interaction with the gluon field~\cite{Faustov1970, Faustov1973}:
\begin{gather}
\Gamma^{\mu}(\mathbf{p}, \mathbf{q}) = \overline{u}_{Q_{1}}(p_{1}) \gamma^{\mu} u_{Q_{1}}(q_{1}) (2\pi)^{3} \delta(\mathbf{p_{2}}-\mathbf{q_{2}}) + \overline{u}_{Q_{2}}(p_{2}) \gamma^{\mu} u_{Q_{2}}(q_{2}) (2\pi)^{3} \delta(\mathbf{p_{1}}-\mathbf{q_{1}}), \label{EqGmu}
\\[5pt]
\begin{cases}
	q_{l} = \varepsilon_{l}(q)\frac{\mathcal{Q}}{M_{d}(q)} \pm \sum\limits_{i=1}^ {3} n^{(i)}(\mathcal{Q})q^{i}, \\
	p_{l} = \varepsilon_{l}(p)\frac{\mathcal{P}}{M_{d}(p)} \pm \sum\limits_{i=1}^ {3} n^{(i)}(\mathcal{P})p^{i}, \\
	n^{(i)\mu}(\mathcal{Q}) = \bigg\{ \frac{\mathcal{Q}^{i}}{M_{d}}, \; \delta_{ij} + \frac{\mathcal{Q}^{i}\mathcal{Q}^{j}}{M_{d}(q)(E_{d}(\mathcal{Q})+M_{d}(q)) }\bigg\}, \\
	M_{d}(q) = \varepsilon_{1}(q)+\varepsilon_{2}(q), 
\end{cases} \qquad l=1,2 \equiv Q, Q'.
\end{gather}
\par
To take into account the finite size of the diquark, it is necessary to calculate the matrix elements of quark currents between diquarks $<d(\mathcal{P})|J_{\mu}|d(\mathcal{Q})>$. These matrix elements are elastic (diagonal) and can be parametrized by the set of form factors $h_{+,1,2,3}(k^{2})$~\cite{U27Bach42misc}.
\noindent
For a scalar diquark:
\begin{equation}
\label{EqMatrS}
<S(\mathcal{P})|J_{\mu}|S(\mathcal{Q})> = h_{+}(k^{2})(\mathcal{P}+\mathcal{Q})_{\mu},
\end{equation}
\noindent
For an axialvector diquark:
\begin{align}
\label{EqMatrA}
<A(\mathcal{P})|J_{\mu}|A(\mathcal{Q})> & = - h_{1}(k^{2}) \bigg[ \epsilon_{d}^{*}(\mathcal{P}) \cdot \epsilon_{d }(\mathcal{Q}) \bigg] (\mathcal{P}+\mathcal{Q})_{\mu} \nonumber
\\
& + h_{2}(k^{2}) \bigg\{ \Big[ \epsilon_{d}^{*}(\mathcal{P}) \cdot \mathcal{Q} \Big] \epsilon_{d; \mu}(\mathcal{Q}) + \Big[ \epsilon_{d}(\mathcal{Q}) \cdot \mathcal{P} \Big] \epsilon_{d; \mu}^{*}(\mathcal{P}) \bigg\} \nonumber
\\
& + h_{3}(k^{2})\frac{1}{M_{A}^{2}} \bigg[ \epsilon_{d}^{*}(\mathcal{P}) \cdot \mathcal{Q} \bigg] \bigg[ \epsilon_{d}(\mathcal{Q}) \cdot \mathcal{P} \bigg] (\mathcal{P}+\mathcal{Q})_{\mu},
\end{align}
where $M_{A}$ is the mass of the axialvector diquark.
\par
\begin{figure}[H]
\center{\includegraphics[width=5.5 cm]{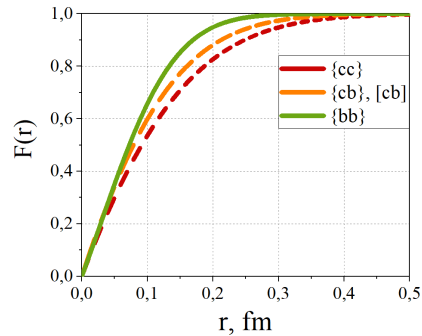}}
\caption{Form factors $F(r)$ for the various doubly heavy diquarks. $\{Q,Q' \}$ denotes axialvector and $[Q,Q']$ denotes scalar diquarks, respectively.}
\label{FormF}
\end{figure}

The calculation shows that~\cite{Baryo2005}:
\begin{equation}
\begin{cases}
	h_{+}(k^{2})=h_{1}(k^{2})=h_{2}(k^{2})=F(\mathbf{k}^{2}), \\
	h_{3}(k^{2})=0, 
\end{cases} \label{Eqh}
\end{equation}
where $F(\mathbf{k}^{2})$ is the form factor in the momentum space:
\begin{align} 
F(\mathbf{k}^{2}) = \frac{\sqrt{M_{d}E_{d}}}{M_{d}+E_{d}} \int \frac{d^{3}p}{(2\pi)^{3}} \Bigg\{ & \bigg[ \overline{\Psi}_{d} \Big( \mathbf{p} + \frac{2\varepsilon_{Q_{2}}(p)}{M_{d}+E_{d}}\mathbf{k} \Big) \sqrt{\frac{\varepsilon_{Q_{1}}(p)+m_{Q_{1}}}{\varepsilon_{Q_{1}}(p+k)+m_{Q_{1}}}} \nonumber
\\
& \hspace{-3.3cm} \times \Big( \frac{\varepsilon_{Q_{1}}(p+k)+\varepsilon_{Q_{1}}(p)}{2\sqrt{\varepsilon_{Q_{1}}(p+k)\varepsilon_{Q_{1}}(p)}} + \frac{\mathbf{pk}}{2(\varepsilon_{Q_{1}}(p)+m_{Q_{1}})\sqrt{\varepsilon_{Q_{1}}(p+k)\varepsilon_{Q_{1}}(p)}} \Big) \Psi_{d}(\mathbf{p}) \bigg] \nonumber 
\\ 
+ & \bigg[ \overline{\Psi}_{d} \Big( \mathbf{p} + \frac{2\varepsilon_{Q_{1}}(p)}{M_{d}+E_{d}}\mathbf{k} \Big) \sqrt{\frac{\varepsilon_{Q_{2}}(p)+m_{Q_{2}}}{\varepsilon_{Q_{2}}(p+k)+m_{Q_{2}}}} \nonumber
\\
& \hspace{-3.3cm} \times \Big( \frac{\varepsilon_{Q_{2}}(p+k)+\varepsilon_{Q_{2}}(p)}{2\sqrt{\varepsilon_{Q_{2}}(p+k)\varepsilon_{Q_{2}}(p)}} + \frac{\mathbf{pk}}{2(\varepsilon_{Q_{2}}(p)+m_{Q_{2}})\sqrt{\varepsilon_{Q_{2}}(p+k)\varepsilon_{Q_{2}}(p)}} \Big) \Psi_{d}(\mathbf{p}) \bigg] \Bigg\}. \label{EqFk}
\end{align} 
The form factor $F(r)$ is determined by the Fourier transform of the $\frac{F(\mathbf{k}^{2})}{\mathbf{k}^{2}}$ which is then multiplied by r. Numerical calculations show that it can be parameterized with high accuracy as~\cite{U27Bach42misc}:
\begin{equation}
\label{EqFr}
F(r) = 1 - e^{-\xi r - \zeta r^{2}},
\end{equation}
the accuracy of this approximation is shown in Fig.~\ref{FormF}.

\par
Finally, we obtain the diquark-antidiquark interaction potential~\cite{Diquark2007, Savch2021}:
\begin{align}
V(r) & = \textcolor{Mahogany}{\pmb{\Bigg[}} V_{\rm Coul.}(r)+V_{\rm conf.}(r) + \frac{1}{E_{1}E_{2}} \Bigg\{ \mathbf{p}\bigg[ V_{\rm Coul.}(r)+V_{\rm conf.}^{V}(r)\bigg] \mathbf{p} \nonumber
\\
& \quad \qquad \qquad \qquad \qquad \qquad \qquad \qquad -\frac{1}{4}\Delta V_{\rm conf.}^{V}(r) + V_{\rm Coul.}^{'}(r)\frac{\mathbf{L}^{2}}{2r} \Bigg\} \textcolor{Mahogany}{\pmb{\Bigg]}_{a}} \label{EqVa} \tag{a}
\\
& + \textcolor{Mahogany}{\pmb{\Bigg[}} \Bigg\{ \frac{1}{2} \bigg[ \frac{1}{E_{1}(E_{1}+M_{1})} + \frac{1}{E_{2}(E_{2}+M_{2})} \bigg] \frac{V_{\rm Coul.}^{'}(r)}{r} \nonumber
\\
& \quad - \frac{1}{2} \bigg[ \frac{1}{M_{1}(E_{1}+M_{1})} + \frac{1}{M_{2}(E_{2}+M_{2})} \bigg] \frac{V_{\rm conf.}^{'}(r)}{r} + \frac{\mu_{d}}{4} \bigg[ \frac{1}{M_{1}^{2}} + \frac{1}{M_{2}^{2}} \bigg] \frac{V_{\rm conf.}^{'V}(r)}{r} \nonumber
\\ 
& \quad + \frac{1}{E_{1}E_{2}} \bigg[ V_{\rm Coul.}^{'}(r) + \frac{\mu_{d}}{4}\Big( \frac{E_{1}}{M_{1}}+\frac{E_{2}}{M_{2}}\Big) V_{\rm conf.}^{'V}(r) \bigg] \frac{1}{r} \Bigg\} \mathbf{L}(\mathbf{S_{1}}+\mathbf{S_{2}}) \nonumber 
\\ 
& + \Bigg\{ \frac{1}{2} \bigg[ \frac{1}{E_{1}(E_{1}+M_{1})} - \frac{1}{E_{2}(E_{2}+M_{2})} \bigg] \frac{V_{\rm Coul.}^{'}(r)}{r} \nonumber
\\
& \quad -\frac{1}{2} \bigg[ \frac{1}{M_{1}(E_{1}+M_{1})} - \frac{1}{M_{2}(E_{2}+M_{2})} \bigg] \frac{V_{\rm conf.}^{'}(r)}{r} + \frac{\mu_{d}}{4} \bigg[ \frac{1}{M_{1}^{2}} - \frac{1}{M_{2}^{2}} \bigg] \frac{V_{\rm conf.}^{'V}(r)}{r} \nonumber
\\
& \quad +\frac{1}{E_{1}E_{2}} \frac{\mu_{d}}{4}\Big( \frac{E_{1}}{M_{1}}-\frac{E_{2}}{M_{2}}\Big) \frac{V_{\rm conf.}^{'V}(r)}{r} \Bigg\} \mathbf{L}(\mathbf{S_{1}}-\mathbf{S_{2}}) \textcolor{Mahogany}{\pmb{\Bigg]}_{b}} \label{EqVb} \tag{b}
\\ 
& + \textcolor{Mahogany}{\pmb{\Bigg[}} \frac{1}{3E_{1}E_{2}} \Bigg\{ \frac{1}{r}V_{\rm Coul.}^{'}(r)-V_{\rm Coul.}^{''}(r) + \frac{\mu_{d}^{2}}{4}\frac{E_{1}E_{2}}{M_{1}M_{2}}\Big( \frac{1}{r}V_{\rm conf.}^{'V}(r)-V_{\rm conf.}^{''V}(r)\Big) \Bigg\} \nonumber 
\\ 
& \qquad \quad \times \bigg[ \frac{3}{r^{2}}\Big( \mathbf{S_{1}r}\Big) \Big(\mathbf{S_{2}r}\Big) -\mathbf{S_{1}S_{2}}\bigg] \textcolor{Mahogany}{\pmb{\Bigg]}_{c}} \label{EqVc} \tag{c}
\\
& + \textcolor{Mahogany}{\pmb{\Bigg[}} \frac{2}{3E_{1}E_{2}} \Bigg\{ \Delta V_{\rm Coul.}(r) + \frac{\mu_{d}^{2}}{4} \frac{E_{1}E_{2}}{M_{1}M_{2}} \Delta V_{\rm conf.}^{V}(r) \Bigg\} \mathbf{S_{1}S_{2}} \textcolor{Mahogany}{\pmb{\Bigg]}_{d}}, \label{EqVd} \tag{d}
\\
\label{EqV}
\end{align}
where $\mathbf{p}$ is the relative momentum, $M_{1, 2}$ and $E_{1, 2}$ are the masses and energies of the diquark and antidiquark, $\mu_{d}$ is the total chromomagnetic moment of the diquark (we chose it to be zero), $\mathbf{S_{d}}$ is the axialvector diquark spin, $\mathbf{L}$ is the relative orbital momentum of the system, $V_{\rm conf.}$ is the confining potential in the nonrelativistic limit:
\begin{equation}
\label{EqConfPot}
V_{\rm conf.}=V_{\rm conf.}^{V}+V_{\rm conf.}^{S}=(1-\varepsilon)(Ar+B) + \varepsilon(Ar+B)=Ar+B,
\end{equation}
where $\varepsilon$ are the scalar and vector confinement mixing coefficient, and the Coulomb potential  $V_{\rm Coul.}(r)$ is taken to be
\begin{equation} 
\label{EqCoul}
V_{\rm Coul.}(r) \equiv -\frac{4}{3}\alpha_{s}\frac{F_{1}(r)F_{2}(r)} {r}.
\end{equation}
$F_{1, 2}(r)$ are the form factors that  take into account diquark sizes (Eq.~(\ref{EqFr})).
\par
In Eq.~(\ref{EqV}) we explicitly separated the spin-independent~(\ref{EqVa}) and spin-dependent terms:~(\ref{EqVb}) for the spin-orbit,~(\ref{EqVc}) for the tensor and~(\ref{EqVd}) for the spin-spin interactions.
\par
First, we calculate the masses and wave functions of the doubly-heavy (anti)diquarks as the bound (anti)quark-(anti)quark states. It is done by solving Eq.~(\ref{EqKv}) with the quasipotential~(\ref{EqPot}), (\ref{EqMatrElqq})-(\ref{EqConf}) numerically. Then the masses of the tetraquarks and their wave functions are obtained for the bound diquark-antidiquark states with the same method. 
\par
Parameters such as the confinement potential mixing coefficient $\varepsilon$, anomalous chromomagnetic moment $\kappa$, parameter of the running coupling constant $\Lambda$, confining potential parameters $A, B$ and quark masses $m_{c,b}$ are taken from our previous works on the study of the properties of mesons and baryons~\cite{Param1986, Param1990, Param1992, Param1995} and are given in the Table~\ref{TabPar1}. The diquark masses $M_{cc, cb, bb}$ and the parameters of their form factors $\xi$ and $\zeta$ have already been calculated earlier~\cite{U27Bach42misc, Diquark2007} and are given in the Table~\ref{TabPar2}.

\begin{table}[H]
\caption{Parameters of the model~\cite{Param1986, Param1990, Param1992, Param1995}.}
\label{TabPar1}
\begin{tabularx}{\textwidth}{>{\rowfont}C>{\rowfont}C>{\rowfont}C>{\rowfont}C>{\rowfont}C>{\rowfont}C>{\rowfont}C<{\unboldrow}}
\toprule
\boldrow $\mathbf{m_{c}}$, GeV & $\mathbf{m_{b}}$, GeV & $\mathbf{A}$, $\text{GeV}^{2}$ & $\mathbf{B}$, GeV & $\boldsymbol\Lambda$, MeV & $\boldsymbol\varepsilon$ & $\boldsymbol\kappa$
\\
\midrule
1.55 & 4.88 & 0.18 & -0.3 & 414 & -1 & -1
\\ 
\bottomrule
\end{tabularx}
\end{table}

\begin{table}[H]
\caption{Masses $M_{QQ'}$ and form factor parameters $\xi, \zeta$ of diquarks. $d$ is the axialvector (A) or scalar (S) diquark. $[Q,Q']$ and $\{Q,Q'\}$ denote combinations of quarks antisymmetric and symmetric in flavor, respectively~\cite{U27Bach42misc, Diquark2007}.}
\label{TabPar2}
\begin{tabularx}{\textwidth}{>{\rowfont}C>{\rowfont}C>{\rowfont}Y{1.8cm}>{\rowfont}C>{\rowfont}C>{\rowfont}Y{1.8cm}>{\rowfont}C>{\rowfont}C<{\unboldrow}}
\toprule
\boldrow \multirow{2}{*}{$\mathbf{QQ'}$} & \multirow{2}{*}{d} & \multicolumn{3}{c}{$\mathbf{Q=c}$} & \multicolumn{3}{c}{$\mathbf{Q=b}$}
\\
\cline{3-8}
\boldrow &  & $\mathbf{M_{cQ}}$, MeV & $\boldsymbol\xi$, GeV & $\boldsymbol\zeta$, GeV$^2$ & $\mathbf{M_{bQ}}$, MeV & $\boldsymbol\xi$, GeV & $\boldsymbol\zeta$, GeV$^2$
\\ 
\midrule
$[Q,c]$ & S &  &  &  & 6519 & 1.50 & 0.59
\\ 
$\{ Q,c \}$ & A & 3226 & 1.30 & 0.42 & 6526 & 1.50 & 0.59
\\
$\{ Q,b \}$ & A & 6526 & 1.50 & 0.59 & 9778 & 1.30 & 1.60
\\
\bottomrule
\end{tabularx}
\end{table}

\section{Masses of fully-heavy tetraquarks}
\label{SecRes}

\par
The calculated mass spectra of fully-heavy tetraquarks are given in Table~\ref{TabRes}. Masses of the ground states (1S) of all possible nine compositions of fully-heavy tetraquarks (including symmetrical: $cc\bar c\bar c, \; cb\bar c\bar b, \; bb\bar b\bar b$, ``mirrored'': $cc\bar b\bar b, \; bb\bar c\bar c$ and nonsymmetrical: $cc\bar c\bar b, cb\bar c\bar c, \; cb\bar b\bar b, \; bb\bar c\bar b$) have already been calculated in our previous work~\cite{Savch2020}.
\par
As it  already have been discussed in Sec.~\ref{SecModel}, a scalar (anti)diquark can be a part of a tetraquark only if the (anti)quarks that form it have different flavors. This means that the $cc\bar c\bar c$ and $bb\bar b\bar b$ tetraquarks can consist only of an axialvector diquarks and antidiquarks, while $cb\bar c\bar b$ can also consist of a scalar and a mixture of axialvector and scalar diquarks and antidiquarks. As the result, we get more possible states for $cb\bar c\bar b$: additional 12 mixed and 6 scalar states are added to 32 axialvector states.

\begin{table}[H] 
\caption{Masses $M_{QQ'\bar Q\bar Q'}$ of the ground (1S) and excited (1P, 2S, 1D, 2P, 3S) $cc\bar c \bar c$, $cb\bar c\bar b$, $bb\bar b\bar b$  states. $d$ and $\bar d'$ are the axialvector ($A$) or scalar ($S$) diquark and antidiquark, respectively. $S$ is the total spin of the diquark-antidiquark system.  All masses are given in MeV.}
\label{TabRes}
\begin{tabularx}{\textwidth}{Y{2.5cm}CCCCCC}
\toprule
$\mathbf{d\bar d'}$ & \textbf{State} & $\mathbf{S}$ & $\mathbf{J^{PC}}$ & $\mathbf{M_{cc\bar c\bar c}}$ & $\mathbf{M_{cb\bar c\bar b}}$ & $\mathbf{M_{bb\bar b\bar b}}$ 
\\
\midrule
\multirow{32}{*}{$A\bar A$} & \multirow{3}{*}{1S} & 0 & $0^{++}$ & 6190 & 12838 & 19315
\\
\cline{3-7}
 & & 1 & $1^{+-}$ & 6271 & 12855 & 19320
\\
\cline{3-7}
 & & 2 & $2^{++}$ & 6367 & 12883 & 19331
\\
\cline{2-7}
 & \multirow{7}{*}{1P} & 0 & $1^{--}$ & 6631 & 13103 & 19536
\\
\cline{3-7}
 & & \multirow{3}{*}{1} & $0^{-+}$ & 6628 & 13100 & 19533
\\
\cline{4-7}
 & & & $1^{-+}$ & 6634 & 13103 & 19535
\\
\cline{4-7}
 & & & $2^{-+}$ & 6644 & 13108 & 19539
\\
\cline{3-7}
 & &\multirow{3}{*}{2} & $1^{--}$ & 6635 & 13103 & 19534
\\
\cline{4-7}
 & & & $2^{--}$ & 6648 & 13109 & 19538
\\
\cline{4-7}
 & & & $3^{--}$ & 6664 & 13116 & 19545
\\
\cline{2-7}
 & \multirow{3}{*}{2S} & 0 & $0^{++}$ & 6782 & 13247 & 19680
\\
\cline{3-7}
 & & 1 & $1^{+-}$ & 6816 & 13256 & 19682
\\
\cline{3-7}
 & & 2 & $2^{++}$ & 6868 & 13272 & 19687
\\
\cline{2-7}
 & \multirow{9}{*}{1D} & 0 & $2^{++}$ & 6921 & 13306 & 19715
\\
\cline{3-7}
 & & \multirow{3}{*}{1} & $1^{+-}$ & 6909 & 13299 & 19710
\\
\cline{4-7}
 & & & $2^{+-}$ & 6920 & 13304 & 19714
\\
\cline{4-7}
 & & & $3^{+-}$ & 6932 & 13311 & 19720
\\
\cline{3-7}
 & & \multirow{5}{*}{2} & $0^{++}$ & 6899 & 13293 & 19705
\\
\cline{4-7}
 & & & $1^{++}$ & 6904 & 13296 & 19707
\\
\cline{4-7}
 & & & $2^{++}$ & 6915 & 13301 & 19711
\\
\cline{4-7}
 & & & $3^{++}$ & 6929 & 13308 & 19717
\\
\cline{4-7}
 & & & $4^{++}$ & 6945 & 13317 & 19724
\\
\cline{2-7}
 & \multirow{7}{*}{2P} & 0 & $1^{--}$ & 7091 & 13428 & 19820
\\
\cline{3-7}
 & & \multirow{3}{*}{1} & $0^{-+}$ & 7100 & 13431 & 19821
\\
\cline{4-7}
 & & & $1^{-+}$ & 7099 & 13431 & 19821
\\
\cline{4-7}
 & & & $2^{-+}$ & 7098 & 13431 & 19822
\\
\cline{3-7}
 & & \multirow{3}{*}{2} & $1^{--}$ & 7113 & 13434 & 19823
\\
\cline{4-7}
 & & & $2^{--}$ & 7113 & 13435 & 19823
\\
\cline{4-7}
 & & & $3^{--}$ & 7112 & 13436 & 19824
\\
\cline{2-7}
 & \multirow{3}{*}{3S} & 0 & $0^{++}$ & 7259 & 13558 & 19941
\\
\cline{3-7}
 & & 1 & $1^{+-}$ & 7287 & 13566 & 19943
\\
\cline{3-7}
 & & 2 & $2^{++}$ & 7333 & 13580 & 19947
\\
\midrule
\multirow{12}{*}{$\dfrac{1}{\sqrt{2}}\Big( A\bar S \pm S\bar A \Big)$} & 1S & \multirow{12}{*}{1} & $1^{+\pm}$ &  & 12863 & 
\\
\cline{2-2}
\cline{4-7}
 & \multirow{3}{*}{1P} &  & $0^{-\pm}$ &  & 13096 &  
\\
\cline{4-7}
 &  &  & $1^{-\pm}$ &  & 13099 & 
 \\
\cline{4-7}
 &  &  & $2^{-\pm}$ &  & 13104 &  
\\
\cline{2-2}
\cline{4-7}
 & 2S &  & $1^{+\pm}$ &  & 13257 &  
\\
\cline{2-2}
\cline{4-7}
 & \multirow{3}{*}{1D} &  & $1^{+\pm}$ &  & 13293 &  
\\
\cline{4-7}
 &  &  & $2^{+\pm}$ &  & 13298 & 
 \\
\cline{4-7}
 &  &  & $3^{+\pm}$ &  & 13305 &  
\\
\cline{2-2}
\cline{4-7}
 & \multirow{3}{*}{2P} &  & $0^{-\pm}$ &  & 13426 &  
\\
\cline{4-7}
 &  &  & $1^{-\pm}$ &  & 13426 & 
\\
\cline{4-7}
 &  &  & $2^{-\pm}$ &  & 13427 &  
\\
\cline{2-2}
\cline{4-7}
 & 3S &  & $1^{+\pm}$ &  & 13566 &  
\\
\midrule
\multirow{6}{*}{$S\bar S$} & 1S & \multirow{6}{*}{0} & $0^{++}$ &  & 12856 & 
\\
\cline{2-2}
\cline{4-7}
 & 1P &  & $1^{--}$ &  & 13095 &  
\\
\cline{2-2}
\cline{4-7}
 & 2S &  & $0^{++}$ &  & 13250 &  
\\
\cline{2-2}
\cline{4-7}
 & 1D &  & $2^{++}$ &  & 13293 &  
\\
\cline{2-2}
\cline{4-7}
 & 2P &  & $1^{--}$ &  & 13420 &  
\\
\cline{2-2}
\cline{4-7}
 & 3S &  & $0^{++}$ &  & 13559 &  
\\
\bottomrule
\end{tabularx}
\end{table}

\section{Threshold analysis}
\label{SecThr}

\par
If a mass of the tetraquark exceeds the sum of the masses of a meson pair composed of the same flavor quarks and antiquarks, and its decay is not forbidden by quantum numbers (spin-parity $J^{PC}$), then the tetraquark will decay into this meson pair through the quark rearrangement via the strong interaction. This is the so-called fall-apart process, which rate is governed by the difference of the tetraquark and threshold masses.  If a mass of the tetraquark lies below the corresponding threshold, the decay is possible due to the heavy quark-antiquark annihilation into gluons or a radiative decay, but such processes are suppressed, making these tetraquarks  narrow states.
\par
In Tables~\ref{TabThrCC}-\ref{TabThrBB} comparisons of mass spectra of fully heavy tetraquarks, calculated by us (Table~\ref{TabRes}), with the meson pair decay thresholds are given. The values of the phase volume $\Delta$ are of special interest:
\begin{equation}
\label{EqDelta}
\Delta = M_{QQ'\bar Q\bar Q'} - M_{thr},
\end{equation}
where $M_{QQ'\bar Q\bar Q'}$ is the tetraquark mass and $M_{thr}$ is the meson pair decay threshold. We are interested in the most probable decay modes for each tetraquark. They, in turn, correspond to the largest of possible values of $\Delta$: $\Delta_{max}$. Therefore, in Tables~\ref{TabThrCC}-\ref{TabThrBB} we compare tetraquark masses not with all possible thresholds, but only with the lowest ones (${[M_{\rm thr}]_{\rm min} \rightarrow \Delta_{\rm max}\to {\rm more \ probable\ decay \ mode}}$).


\small
\setlength\LTleft{-\extralength}
\begin{longtable}{Y{1.36cm}Y{1.2cm}Y{1.2cm}Y{1.2cm}Y{1.2cm}Y{2.0cm}Y{2.0cm}Y{2.0cm}Y{2.5cm}} 
\caption{Masses $M$ of the ground (1S) and excited (1P, 2S, 1D, 2P, 3S) $cc\bar c \bar c$ states composed from the axialvector diquarks (Table~\ref{TabRes}) and the corresponding meson-meson thresholds. $d$ and $\bar d'$ are the axialvector (A) or scalar (S) diquark and antidiquark, respectively.  $S$ is the total spin of the diquark-antidiquark system. $M_{thr}$ is the corresponding meson-meson threshold~\cite{PDG2022}. $\Delta$ is the difference between the tetraquark mass and threshold: $\Delta = M - M_{thr}$. All masses are given in MeV. For the states with the maximum $\Delta$ (corresponding to lightest threshold, main decay channel) less than 300 MeV, all possible thresholds and their $\Delta$ are given. For the states with maximum $\Delta$ above 300 MeV, only the lightest  thresholds are shown. The states with maximum $\Delta$  less than 100 MeV are additionally highlighted in violet as most promising to be stable. The states with negative maximum $\Delta$  are highlighted in red for the same reason. We also give thresholds with a small negative $\Delta$, since we do not take into account the errors of theoretical calculations. The candidates for the states recently observed  by LHCb~\cite{U10-ccLHCb2020}, CMS~\cite{ccCMS2022} and ATLAS~\cite{ccATLAS2022} are highlighted in color: turquoise for X(6200) (ATLAS), emerald for X(6400) (LHCb) and X(6600) (CMS, ATLAS), green for X(6900) (LHCb, CMS, ATLAS), blue for X(7200) (LHCb, ATLAS) and X(7300) (CMS). Additionally all di-$J/\psi$ and di-$\Upsilon(1S)$ (in similar table for $bb\bar b\bar b$ states) thresholds are shown in bold since this meson pairs are easiest to study in experiments.}
\label{TabThrCC}
\\
\toprule
$\mathbf{QQ\bar Q\bar Q}$ & $\mathbf{d\bar d'}$ & \textbf{State} & $\mathbf{S}$ & $\mathbf{J^{PC}}$ & $\mathbf{M}$ & $\mathbf{M_{thr}}$ & $\boldsymbol\Delta$ & \textbf{Meson pair}
\\
\midrule
\endfirsthead
\multicolumn{9}{c}{\hspace{-\extralength}\textbf{Table} \thetable . Table continued.}
\\
\toprule
$\mathbf{QQ\bar Q\bar Q}$ & $\mathbf{d\bar d'}$ & \textbf{State} & $\mathbf{S}$ & $\mathbf{J^{PC}}$ & $\mathbf{M}$ & $\mathbf{M_{thr}}$ & $\boldsymbol\Delta$ & \textbf{Meson pair}
\\
\midrule
\endhead
\multirow{21}{*}{$cc\bar c\bar c$} & \multirow{21}{*}{$A\bar A$} & \multirow{4}{*}{1S} & \multirow{2}{*}{0} & \multirow{2}{*}{\textcolor{Aquamarine}{$\mathbf{0^{++}}$}} & \multirow{2}{*}{\textcolor{Aquamarine}{$\mathbf{6190}$}} & 5968 & \textcolor{black}{222} & $\eta_{c}(1S)\eta_{c}(1S)$ 
\\
\cline{7-9}
 & & & & & & $\mathbf{6194}$ & \textcolor{Aquamarine}{$\mathbf{-4}$} & \boldmath $J/\psi(1S)J/\psi(1S)$
\\
\cline{4-9}
 & & & 1 & \textcolor{black}{$1^{+-}$} & \textcolor{black}{6271} & 6081 & \textcolor{black}{190} & $\eta_{c}(1S)J/\psi(1S)$
\\
\cline{4-9}
 & & & 2 & \textcolor{ForestGreen}{$\mathbf{2^{++}}$} & \textcolor{ForestGreen}{$\mathbf{6367}$} & $\mathbf{6194}$ & \textcolor{ForestGreen}{$\mathbf{173}$} & \boldmath $J/\psi(1S)J/\psi(1S)$
\\
\cline{3-9}
 & & \multirow{15}{*}{1P} & \multirow{3}{*}{0} & \multirow{3}{*}{\textcolor{black}{$1^{--}$}} & \multirow{3}{*}{\textcolor{black}{6631}} & 6509 & \textcolor{black}{122} & $\eta_{c}(1S)h_{c}(1P)$
 \\
\cline{7-9}
 & & & & & & 6512 & 119 & $J/\psi(1S)\chi_{c0}(1P)$
\\
\cline{7-9}
 & & & & & & 6608 & 23 & $J/\psi(1S)\chi_{c1}(1P)$
\\
\cline{4-9}
 & & & \multirow{6}{*}{1} & \multirow{2}{*}{\textcolor{black}{$0^{-+}$}} & \multirow{2}{*}{\textcolor{black}{6628}} & 6399 & \textcolor{black}{229} & $\eta_{c}(1S)\chi_{c0}(1P)$
\\
\cline{7-9}
 & & & & & & 6622 & 6 & $J/\psi(1S)h_{c}(1P)$
\\
\cline{5-9}
 & & & & \multirow{2}{*}{\textcolor{black}{$1^{-+}$}} & \multirow{2}{*}{\textcolor{black}{6634}} & 6495 & \textcolor{black}{139} & $\eta_{c}(1S)\chi_{c1}(1P)$
\\
\cline{7-9}
 & & & & & & 6622 & 12 & $J/\psi(1S)h_{c}(1P)$
\\
\cline{5-9}
 & & & & \multirow{2}{*}{\textcolor{black}{$2^{-+}$}} & \multirow{2}{*}{\textcolor{black}{6644}} & 6540 & \textcolor{black}{104} & $\eta_{c}(1S)\chi_{c2}(1P)$
\\
\cline{7-9}
 & & & & & & 6622 & 22 & $J/\psi(1S)h_{c}(1P)$
\\
\cline{4-9}
 & & & \multirow{7}{*}{2} & \multirow{4}{*}{\textcolor{black}{$1^{--}$}} & \multirow{4}{*}{\textcolor{black}{6635}} & 6509 & \textcolor{black}{126} & $\eta_{c}(1S)h_{c}(1P)$
\\
\cline{7-9}
 & & & & & & 6512 & 123 & $J/\psi(1S)\chi_{c0}(1P)$
\\
\cline{7-9}
 & & & & & & 6608 & 27 & $J/\psi(1S)\chi_{c1}(1P)$
 \\
\cline{7-9}
 & & & & & & 6653 & -18 & $J/\psi(1S)\chi_{c2}(1P)$
\\
\cline{5-9}
 & & & & \multirow{2}{*}{\textcolor{RedViolet}{$2^{--}$}} & \multirow{2}{*}{\textcolor{RedViolet}{6648}} & 6608 & \textcolor{RedViolet}{40} & $J/\psi(1S)\chi_{c1}(1P)$
\\
\cline{7-9}
 & & & & & & 6653 & \textcolor{RedViolet}{-5} & $J/\psi(1S)\chi_{c2}(1P)$
\\
\cline{5-9}
 & & & & \textcolor{RedViolet}{$3^{--}$} & \textcolor{RedViolet}{6664} & 6653 & \textcolor{RedViolet}{11} & $J/\psi(1S)\chi_{c2}(1P)$
\\
\cline{3-9}
\pagebreak
\multirow{25}{*}{$cc\bar c\bar c$} & \multirow{25}{*}{$A\bar A$} & \multirow{4}{*}{2S} & \multirow{2}{*}{0} & \multirow{2}{*}{\textcolor{ForestGreen}{$\mathbf{0^{++}}$}} & \multirow{2}{*}{\textcolor{ForestGreen}{$\mathbf{6782}$}\textsuperscript{1}} & 5968 & 814 & $\eta_{c}(1S)\eta_{c}(1S)$
\\
\cline{7-9}
 & & & & & & $\mathbf{6194}$ & \textcolor{ForestGreen}{$\mathbf{588}$} & \boldmath $J/\psi(1S)J/\psi(1S)$
\\
\cline{4-9}
 & & & 1 & $1^{+-}$ & 6816 & 6081 & 735 & $\eta_{c}(1S)J/\psi(1S)$
\\
\cline{4-9}
 & & & 2 & \textcolor{LimeGreen}{$\mathbf{2^{++}}$} & \textcolor{LimeGreen}{$\mathbf{6868}$} & $\mathbf{6194}$ & \textcolor{LimeGreen}{$\mathbf{674}$} & \boldmath $J/\psi(1S)J/\psi(1S)$
\\
\cline{3-9}
 & & \multirow{11}{*}{1D} & 0 & \textcolor{LimeGreen}{$\mathbf{2^{++}}$} & \textcolor{LimeGreen}{$\mathbf{6921}$} & $\mathbf{6194}$ & \textcolor{LimeGreen}{$\mathbf{727}$} & \boldmath $J/\psi(1S)J/\psi(1S)$
\\
\cline{4-9}
 & & & \multirow{3}{*}{1} & $1^{+-}$ & 6909 & 6081 & 828 & $\eta_{c}(1S)J/\psi(1S)$
\\
\cline{5-9}
 & & & & \textcolor{black}{$2^{+-}$} & \textcolor{black}{6920} & 6808 & \textcolor{black}{112} & $\eta_{c}(1S)\psi_{2}(3823)$
\\
\cline{5-9}
 & & & & \textcolor{black}{$3^{+-}$} & \textcolor{black}{6932} & 6827 & \textcolor{black}{105} & $\eta_{c}(1S)\psi_{3}(3842)$
\\
\cline{4-9}
 & & & \multirow{7}{*}{2} & \multirow{2}{*}{\textcolor{LimeGreen}{$\mathbf{0^{++}}$}} & \multirow{2}{*}{\textcolor{LimeGreen}{$\mathbf{6899}$}} & 5968 & 931 & $\eta_{c}(1S)\eta_{c}(1S)$
\\
\cline{7-9}
 & & & & & & $\mathbf{6194}$ & \textcolor{LimeGreen}{$\mathbf{705}$} & \boldmath $J/\psi(1S)J/\psi(1S)$
\\
\cline{5-9}
 & & & & \textcolor{LimeGreen}{$\mathbf{1^{++}}$} & \textcolor{LimeGreen}{$\mathbf{6904}$} & $\mathbf{6194}$ & \textcolor{LimeGreen}{$\mathbf{710}$} & \boldmath $J/\psi(1S)J/\psi(1S)$
\\
\cline{5-9}
 & & & & \textcolor{LimeGreen}{$\mathbf{2^{++}}$} & \textcolor{LimeGreen}{$\mathbf{6915}$} & $\mathbf{6194}$ & \textcolor{LimeGreen}{$\mathbf{721}$} & \boldmath $J/\psi(1S)J/\psi(1S)$
\\
\cline{5-9}
 & & & & \multirow{2}{*}{\textcolor{RedViolet}{$3^{++}$}} & \multirow{2}{*}{\textcolor{RedViolet}{6929}} & 6921 & \textcolor{RedViolet}{8} & $J/\psi(1S)\psi_{2}(3823)$
\\
\cline{7-9}
 & & & & & & 6940 & \textcolor{RedViolet}{19} & $J/\psi(1S)\psi_{3}(3842)$
\\
\cline{5-9}
 & & & & \textcolor{RedViolet}{$4^{++}$} & \textcolor{RedViolet}{6945} & 6940 & \textcolor{RedViolet}{5} & $J/\psi(1S)\psi_{3}(3842)$
\\
\cline{3-9}
 & & \multirow{7}{*}{2P} & 0 & $1^{--}$ & 7091 & 6509 & 582 & $\eta_{c}(1S)h_{c}(1P)$
\\
\cline{4-9}
 & & & \multirow{3}{*}{1} & $0^{-+}$ & 7100 & 6399 & 701 & $\eta_{c}(1S)\chi_{c0}(1P)$
\\
\cline{5-9}
 & & & & $1^{-+}$ & 7099 & 6495 & 604 & $\eta_{c}(1S)\chi_{c1}(1P)$
\\
\cline{5-9}
 & & & & $2^{-+}$ & 7098 & 6540 & 558 & $\eta_{c}(1S)\chi_{c2}(1P)$
\\
\cline{4-9}
 & & & \multirow{3}{*}{2} & $1^{--}$ & 7113 & 6509 & 604 & $\eta_{c}(1S)h_{c}(1P)$
\\
\cline{5-9}
 & & & & $2^{--}$ & 7113 & 6608 & 505 & $J/\psi(1S)\chi_{c1}(1P)$
\\
\cline{5-9}
 & & & & $3^{--}$ & 7112 & 6653 & 459 & $J/\psi(1S)\chi_{c2}(1P)$
\\
\cline{3-9}
 & & \multirow{4}{*}{3S} & \multirow{2}{*}{0} & \multirow{2}{*}{\textcolor{RoyalBlue}{$\mathbf{0^{++}}$}} & \multirow{2}{*}{\textcolor{RoyalBlue}{$\mathbf{7259}$}} & 5968 & 1291 & $\eta_{c}(1S)\eta_{c}(1S)$
\\
\cline{7-9}
 & & & & & & $\mathbf{6194}$ & \textcolor{RoyalBlue}{$\mathbf{1065}$} & \boldmath $J/\psi(1S)J/\psi(1S)$
\\
\cline{4-9}
 & & & 1 & $1^{+-}$ & 7287 & 6081 & 1206 & $\eta_{c}(1S)J/\psi(1S)$
\\
\cline{4-9}
 & & & 2 & \textcolor{RoyalBlue}{$\mathbf{2^{++}}$} & \textcolor{RoyalBlue}{$\mathbf{7333}$}\textsuperscript{2} & $\mathbf{6194}$ & \textcolor{RoyalBlue}{$\mathbf{1139}$} & \boldmath $J/\psi(1S)J/\psi(1S)$
\\
\bottomrule
 & & & \multicolumn{3}{c}{\footnotesize{\textsuperscript{1} Candidate only for the X(6600) state.}} & & &
\\
 & & & \multicolumn{3}{c}{\footnotesize{\textsuperscript{2} Candidate only for the X(7300) state.}} & & &
\\
\end{longtable}



\small
\setlength\LTleft{-\extralength}
\begin{longtable}{Y{1.36cm}Y{1.2cm}Y{1.2cm}Y{1.2cm}Y{1.2cm}Y{2.0cm}Y{2.0cm}Y{2.0cm}Y{2.5cm}} 
\caption{Same as in Table~\ref{TabThrCC} but for $cb\bar c\bar b$ states composed from the axialvector (A) diquarks.}
\label{TabThrCB1}
\\
\toprule
$\mathbf{QQ\bar Q\bar Q}$ & $\mathbf{d\bar d'}$ & \textbf{State} & $\mathbf{S}$ & $\mathbf{J^{PC}}$ & $\mathbf{M}$ & $\mathbf{M_{thr}}$ & $\boldsymbol\Delta$ & \textbf{Meson pair}
\\
\midrule
\endfirsthead
\multicolumn{9}{c}{\hspace{-\extralength}\textbf{Table} \thetable . Table continued.}
\\
\toprule
$\mathbf{QQ\bar Q\bar Q}$ & $\mathbf{d\bar d'}$ & \textbf{State} & $\mathbf{S}$ & $\mathbf{J^{PC}}$ & $\mathbf{M}$ & $\mathbf{M_{thr}}$ & $\boldsymbol\Delta$ & \textbf{Meson pair}
\\
\midrule
\endhead
\multirow{23}{*}{$cb\bar c\bar b$} & \multirow{23}{*}{$A\bar A$} & \multirow{3}{*}{1S} & 0 & $0^{++}$ & 12838 & 12383 & 455 & $\eta_{c}(1S)\eta_{b}(1S)$ 
\\
\cline{4-9}
 & & & 1 & $1^{+-}$ & 12855 & 12444 & 411 & $\eta_{c}(1S)\Upsilon(1S)$
\\
\cline{4-9}
 & & & 2 & $2^{++}$ & 12883 & 12557 & 326 & $J/\psi(1S)\Upsilon(1S)$
\\
\cline{3-9}
 & & \multirow{20}{*}{1P} & \multirow{8}{*}{0} & \multirow{8}{*}{\textcolor{black}{$1^{--}$}} & \multirow{8}{*}{\textcolor{black}{13103}} & 12875 & \textcolor{black}{228} & $\chi_{c0}(1P)\Upsilon(1S)$ 
  \\
\cline{7-9}
 & & & & & & 12883 & 220 & $\eta_{c}(1S)h_{b}(1P)$
 \\
\cline{7-9}
 & & & & & & 12924 & 179 & $h_{c}(1P)\eta_{b}(1S)$
 \\
\cline{7-9}
 & & & & & & 12956 & 147 & $J/\psi(1S)\chi_{b0}(1P)$
 \\
\cline{7-9}
 & & & & & & 12971 & 132 & $\chi_{c1}(1P)\Upsilon(1S)$
 \\
\cline{7-9}
 & & & & & & 12990 & 113 & $J/\psi(1S)\chi_{b1}(1P)$
 \\
\cline{7-9}
 & & & & & & 13009 & 94 & $J/\psi(1S)\chi_{b2}(1P)$
 \\
\cline{7-9}
 & & & & & & 13016 & 87 & $\chi_{c2}(1P)\Upsilon(1S)$
\\
\cline{4-9}
 & & & \multirow{12}{*}{1} & \multirow{4}{*}{\textcolor{black}{$0^{-+}$}} & \multirow{4}{*}{\textcolor{black}{13100}} & 12813 & \textcolor{black}{287} & $\chi_{c0}(1P)\eta_{b}(1S)$
 \\
\cline{7-9}
 & & & & & & 12843 & 257 & $\eta_{c}(1S)\chi_{b0}(1P)$
 \\
\cline{7-9}
 & & & & & & 12986 & 114 & $h_{c}(1P)\Upsilon(1S)$
 \\
\cline{7-9}
 & & & & & & 12996 & 104 & $J/\psi(1S)h_{b}(1P)$
\\
\cline{5-9}
 & & & & \multirow{4}{*}{\textcolor{black}{$1^{-+}$}} & \multirow{4}{*}{\textcolor{black}{13103}} & 12877 & \textcolor{black}{226} & $\eta_{c}(1S)\chi_{b1}(1P)$
\\
\cline{7-9}
 & & & & & & 12909 & 194 & $\chi_{c1}(1P)\eta_{b}(1S)$
\\
\cline{7-9}
 & & & & & & 12986 & 117 & $h_{c}(1P)\Upsilon(1S)$
\\
\cline{7-9}
 & & & & & & 12996 & 107 & $J/\psi(1S)h_{b}(1P)$
\\
\cline{5-9}
 & & & & \multirow{4}{*}{\textcolor{black}{$2^{-+}$}} & \multirow{4}{*}{\textcolor{black}{13108}} & 12896 & \textcolor{black}{212} & $\eta_{c}(1S)\chi_{b2}(1P)$
\\
\cline{7-9}
 & & & & & & 12955 & 153 & $\chi_{c2}(1P)\eta_{b}(1S)$
\\
\cline{7-9}
 & & & & & & 12986 & 122 & $h_{c}(1P)\Upsilon(1S)$
\\
\cline{7-9}
 & & & & & & 12996 & 112 & $J/\psi(1S)h_{b}(1P)$
\\
\cline{4-9}
\pagebreak
\multirow{39}{*}{$cb\bar c\bar b$} & \multirow{39}{*}{$A\bar A$} & \multirow{14}{*}{1P} & \multirow{14}{*}{2} & \multirow{8}{*}{\textcolor{black}{$1^{--}$}} & \multirow{8}{*}{\textcolor{black}{13103}} & 12875 & \textcolor{black}{228} & $\chi_{c0}(1P)\Upsilon(1S)$
  \\
\cline{7-9}
 & & & & & & 12883 & 220 & $\eta_{c}(1S)h_{b}(1P)$
 \\
\cline{7-9}
 & & & & & & 12924 & 179 & $h_{c}(1P)\eta_{b}(1S)$
 \\
\cline{7-9}
 & & & & & & 12956 & 147 & $J/\psi(1S)\chi_{b0}(1P)$
 \\
\cline{7-9}
 & & & & & & 12971 & 132 & $\chi_{c1}(1P)\Upsilon(1S)$
 \\
\cline{7-9}
 & & & & & & 12990 & 113 & $J/\psi(1S)\chi_{b1}(1P)$
 \\
\cline{7-9}
 & & & & & & 13009 & 94 & $J/\psi(1S)\chi_{b2}(1P)$
 \\
\cline{7-9}
 & & & & & & 13016 & 87 & $\chi_{c2}(1P)\Upsilon(1S)$
\\
\cline{5-9}
 & & & & \multirow{4}{*}{\textcolor{black}{$2^{--}$}} & \multirow{4}{*}{\textcolor{black}{13109}} & 12971 & \textcolor{black}{138} & $\chi_{c1}(1P)\Upsilon(1S)$
 \\
\cline{7-9}
 & & & & & & 12990 & 119 & $J/\psi(1S)\chi_{b1}(1P)$
 \\
\cline{7-9}
 & & & & & & 13009 & 100 & $J/\psi(1S)\chi_{b2}(1P)$
 \\
\cline{7-9}
 & & & & & & 13016 & 93 & $\chi_{c2}(1P)\Upsilon(1S)$
 \\
\cline{5-9}
 & & & & \multirow{2}{*}{\textcolor{black}{$3^{--}$}} & \multirow{2}{*}{\textcolor{black}{13116}} & 13009 & \textcolor{black}{107} & $J/\psi(1S)\chi_{b2}(1P)$
  \\
\cline{7-9}
 & & & & & & 13016 & 100 & $\chi_{c2}(1P)\Upsilon(1S)$
\\
\cline{3-9}
 &  & \multirow{3}{*}{2S} & 0 & $0^{++}$ & 13247 & 12383 & 864 & $\eta_{c}(1S)\eta_{b}(1S)$
\\
\cline{4-9}
 & & & 1 & $1^{+-}$ & 13256 & 12444 & 812 & $\eta_{c}(1S)\Upsilon(1S)$
\\
\cline{4-9}
 & & & 2 & $2^{++}$ & 13272 & 12557 & 715 & $J/\psi(1S)\Upsilon(1S)$
\\
\cline{3-9}
 & & \multirow{12}{*}{1D} & 0 & $2^{++}$ & 13306 & 12557 & 749 & $J/\psi(1S)\Upsilon(1S)$
\\
\cline{4-9}
 & & & \multirow{4}{*}{1} & $1^{+-}$ & 13299 & 12444 & 855 & $\eta_{c}(1S)\Upsilon(1S)$
\\
\cline{5-9}
 & & & & \multirow{2}{*}{\textcolor{black}{$2^{+-}$}} & \multirow{2}{*}{\textcolor{black}{13304}} & 13148 & \textcolor{black}{156} & $\eta_{c}(1S)\Upsilon_{2}(1D)$
\\
\cline{7-9}
 & & & & & & 13222 & 82 & $\psi_{2}(3823)\eta_{b}(1S)$
\\
\cline{5-9}
 & & & & \textcolor{RedViolet}{$3^{+-}$} & \textcolor{RedViolet}{13311} & 13241 & \textcolor{RedViolet}{70} & $\psi_{3}(3842)\eta_{b}(1S)$
\\
\cline{4-9}
 & & & \multirow{7}{*}{2} & $0^{++}$ & 13293 & 12383 & 910 & $\eta_{c}(1S)\eta_{b}(1S)$
\\
\cline{5-9}
 & & & & $1^{++}$ & 13296 & 12557 & 739 & $J/\psi(1S)\Upsilon(1S)$
\\
\cline{5-9}
 & & & & $2^{++}$ & 13301 & 12557 & 744 & $J/\psi(1S)\Upsilon(1S)$
\\
\cline{5-9}
 & & & & \multirow{3}{*}{\textcolor{RedViolet}{$3^{++}$}} & \multirow{3}{*}{\textcolor{RedViolet}{13308}} & 13261 & \textcolor{RedViolet}{47} & $J/\psi(1S)\Upsilon_{2}(1D)$
 \\
\cline{7-9}
 & & & & & & 13284 & \textcolor{RedViolet}{24} & $\psi_{2}(3823)\Upsilon(1S)$
 \\
\cline{7-9}
 & & & & & & 13303 & \textcolor{RedViolet}{5} & $\psi_{3}(3842)\Upsilon(1S)$
\\
\cline{5-9}
 & & & & \textcolor{RedViolet}{$4^{++}$} & \textcolor{RedViolet}{13317} & 13303 & \textcolor{RedViolet}{14} & $\psi_{3}(3842)\Upsilon(1S)$
\\
\cline{3-9}
 & & \multirow{7}{*}{2P} & 0 & $1^{--}$ & 13428 & 12875 & 553 & $\chi_{c0}(1P)\Upsilon(1S)$
\\
\cline{4-9}
 & & & \multirow{3}{*}{1} & $0^{-+}$ & 13431 & 12813 & 618 & $\chi_{c0}(1P)\eta_{b}(1S)$
\\
\cline{5-9}
 & & & & $1^{-+}$ & 13431 & 12877 & 554 & $\eta_{c}(1S)\chi_{b1}(1P)$
\\
\cline{5-9}
 & & & & $2^{-+}$ & 13431 & 12896 & 535 & $\eta_{c}(1S)\chi_{b2}(1P)$
\\
\cline{4-9}
 & & & \multirow{3}{*}{2} & $1^{--}$ & 13434 & 12875 & 559 & $\chi_{c0}(1P)\Upsilon(1S)$
\\
\cline{5-9}
 & & & & $2^{--}$ & 13435 & 12971 & 464 & $\chi_{c1}(1P)\Upsilon(1S)$
\\
\cline{5-9}
 & & & & $3^{--}$ & 13436 & 13009 & 427 & $J/\psi(1S)\chi_{b2}(1P)$
\\
\cline{3-9}
 & & \multirow{3}{*}{3S} & 0 & $0^{++}$ & 13558 & 12383 & 1175 & $\eta_{c}(1S)\eta_{b}(1S)$
\\
\cline{4-9}
 & & & 1 & $1^{+-}$ & 13566 & 12444 & 1122 & $\eta_{c}(1S)\Upsilon(1S)$
\\
\cline{4-9}
 & & & 2 & $2^{++}$ & 13580 & 12557 & 1023 & $J/\psi(1S)\Upsilon(1S)$
\\
\bottomrule
\end{longtable}


\begin{table}
\caption{Same as in Table~\ref{TabThrCC} but for the $cb\bar c\bar b$ states composed from the mixture of axialvector (A) and scalar (S) diquarks.}
\label{TabThrCB2}
\begin{adjustwidth}{-\extralength}{0cm}
\begin{tabularx}{\fulllength}{CY{2.5cm}CCCCCCY{2.5cm}}
\toprule
$\mathbf{QQ\bar Q\bar Q}$ & $\mathbf{d\bar d'}$ & \textbf{State} & $\mathbf{S}$ & $\mathbf{J^{PC}}$ & $\mathbf{M}$ & $\mathbf{M_{thr}}$ & $\boldsymbol\Delta$ & \textbf{Meson pair}
\\
\midrule
\multirow{47}{*}{$cb\bar c\bar b$} & \multirow{47}{*}{$\dfrac{1}{\sqrt{2}}\Big( A\bar S \pm S\bar A \Big)$} & \multirow{2}{*}{1S} & \multirow{47}{*}{1} & $1^{++}$ & \multirow{2}{*}{12863} & 12557 & 306 & $J/\psi(1S)\Upsilon(1S)$
\\
\cline{5-5}
\cline{7-9}
 & & & & $1^{+-}$ & & 12444 & 419 & $\eta_{c}(1S)\Upsilon(1S)$
\\
\cline{3-3}
\cline{5-9}
 & & \multirow{26}{*}{1P} & & \multirow{4}{*}{\textcolor{black}{$0^{-+}$}} & \multirow{6}{*}{\textcolor{black}{13096}} & 12813 & \textcolor{black}{283} & $\chi_{c0}(1P)\eta_{b}(1S)$
\\
\cline{7-9}
 & & & & & & 12843 & 253 & $\eta_{c}(1S)\chi_{b0}(1P)$
\\
\cline{7-9}
 & & & & & & 12986 & 110 & $h_{c}(1P)\Upsilon(1S)$
\\
\cline{7-9}
 & & & & & & 12996 & 100 & $J/\psi(1S)h_{b}(1P)$
\\
\cline{5-5}
\cline{7-9}
 & & & & \multirow{2}{*}{\textcolor{black}{$0^{--}$}} & & 12971 & \textcolor{black}{125} & $\chi_{c1}(1P)\Upsilon(1S)$
\\
\cline{7-9}
 & & & & & & 12990 & 106 & $J/\psi(1S)\chi_{b1}(1P)$
\\
\cline{5-9}
 & & & & \multirow{4}{*}{\textcolor{black}{$1^{-+}$}} & \multirow{12}{*}{\textcolor{black}{13099}} & 12877 & \textcolor{black}{222} & $\eta_{c}(1S)\chi_{b1}(1P)$
\\
\cline{7-9}
 & & & & & & 12909 & 190 & $\chi_{c1}(1P)\eta_{b}(1S)$
\\
\cline{7-9}
 & & & & & & 12986 & 113 & $h_{c}(1P)\Upsilon(1S)$
\\
\cline{7-9}
 & & & & & & 12996 & 103 & $J/\psi(1S)h_{b}(1P)$
\\
\cline{5-5}
\cline{7-9}
 & & & & \multirow{8}{*}{\textcolor{black}{$1^{--}$}} & & 12875 & \textcolor{black}{224} & $\chi_{c0}(1P)\Upsilon(1S)$
\\
\cline{7-9}
 & & & & & & 12883 & 216 & $\eta_{c}(1S)h_{b}(1P)$ 
\\
\cline{7-9}
 & & & & & & 12924 & 175 & $h_{c}(1P)\eta_{b}(1S)$ 
\\
\cline{7-9}
 & & & & & & 12956 & 143 & $J/\psi(1S)\chi_{b0}(1P)$
 \\
\cline{7-9}
 & & & & & & 12971 & 128 & $\chi_{c1}(1P)\Upsilon(1S)$
 \\
\cline{7-9}
 & & & & & & 12990 & 109 & $J/\psi(1S)\chi_{b1}(1P)$
 \\
\cline{7-9}
 & & & & & & 13009 & 90 & $J/\psi(1S)\chi_{b2}(1P)$
 \\
\cline{7-9}
 & & & & & & 13016 & 83 & $\chi_{c2}(1P)\Upsilon(1S)$  
\\
\cline{5-9}
 & & & & \multirow{4}{*}{\textcolor{black}{$2^{-+}$}} & \multirow{8}{*}{\textcolor{black}{13104}} & 12896 & \textcolor{black}{208} & $\eta_{c}(1S)\chi_{b2}(1P)$
\\
\cline{7-9}
 & & & & & & 12955 & 149 & $\chi_{c2}(1P)\eta_{b}(1S)$
\\
\cline{7-9}
 & & & & & & 12986 & 118 & $h_{c}(1P)\Upsilon(1S)$
\\
\cline{7-9}
 & & & & & & 12996 & 108 & $J/\psi(1S)h_{b}(1P)$
\\
\cline{5-5}
\cline{7-9}
 & & & & \multirow{4}{*}{\textcolor{black}{$2^{--}$}} & & 12971 & \textcolor{black}{133} & $\chi_{c1}(1P)\Upsilon(1S)$
\\
\cline{7-9}
 & & & & & & 12990 & 114 & $J/\psi(1S)\chi_{b1}(1P)$ 
\\
\cline{7-9}
 & & & & & & 13009 & 114 & $J/\psi(1S)\chi_{b2}(1P)$
\\
\cline{7-9}
 & & & & & & 13016 & 114 & $\chi_{c2}(1P)\Upsilon(1S)$ 
\\
\cline{3-3}
\cline{5-9}
 & & \multirow{2}{*}{2S} & & $1^{++}$ & \multirow{2}{*}{13257} & 12557 & 700 & $J/\psi(1S)\Upsilon(1S)$
\\
\cline{5-5}
\cline{7-9}
 & & & & $1^{+-}$ & & 12444 & 813 & $\eta_{c}(1S)\Upsilon(1S)$
\\
\cline{3-3}
\cline{5-9}
 & & \multirow{9}{*}{1D} & & $1^{++}$ & \multirow{2}{*}{13293} & 12557 & 736 & $J/\psi(1S)\Upsilon(1S)$
\\
\cline{5-5}
\cline{7-9}
 & & & & $1^{+-}$ & & 12444 & 849 & $\eta_{c}(1S)\Upsilon(1S)$
\\
\cline{5-9}
 & & & & $2^{++}$ & \multirow{3}{*}{\textcolor{black}{13298}} & 12557 & 741 & $J/\psi(1S)\Upsilon(1S)$
\\
\cline{5-5}
\cline{7-9}
 & & & & \multirow{2}{*}{\textcolor{black}{$2^{+-}$}} & & 13148 & \textcolor{black}{150} & $\eta_{c}(1S)\Upsilon_{2}(1D)$
\\
\cline{7-9}
 & & & & & & 13222 & 76 & $\psi_{2}(3823)\eta_{b}(1S)$
\\
\cline{5-9}
 & & & & \multirow{3}{*}{\textcolor{RedViolet}{$3^{++}$}} & \multirow{4}{*}{\textcolor{RedViolet}{13305}} & 13261 & \textcolor{RedViolet}{44} & $J/\psi(1S)\Upsilon_{2}(1D)$
\\
\cline{7-9}
 & & & & & & 13284 & \textcolor{RedViolet}{21} & $\psi_{2}(3823)\Upsilon(1S)$
\\
\cline{7-9}
 & & & & & & 13303 & \textcolor{RedViolet}{2} & $\psi_{3}(3842)\Upsilon(1S)$
\\
\cline{5-5}
\cline{7-9}
 & & & & \textcolor{RedViolet}{$3^{+-}$} & & 13241 & \textcolor{RedViolet}{64} & $\psi_{3}(3842)\eta_{b}(1S)$
\\
\cline{3-3}
\cline{5-9}
 & & \multirow{6}{*}{2P} & & $0^{-+}$ & \multirow{2}{*}{13426} & 12813 & 613 & $\chi_{c0}(1P)\eta_{b}(1S)$
\\
\cline{5-5}
\cline{7-9}
 & & & & $0^{--}$ & & 12971 & 455 & $\chi_{c1}(1P)\Upsilon(1S)$
\\
\cline{5-9}
 & & & & $1^{-+}$ & \multirow{2}{*}{13426} & 12877 & 549 & $\eta_{c}(1S)\chi_{b1}(1P)$
\\
\cline{5-5}
\cline{7-9}
 & & & & $1^{--}$ & & 12875 & 551 & $\chi_{c0}(1P)\Upsilon(1S)$
\\
\cline{5-9}
 & & & & $2^{-+}$ & \multirow{2}{*}{13427} & 12896 & 531 & $\eta_{c}(1S)\chi_{b2}(1P)$
\\
\cline{5-5}
\cline{7-9}
 & & & & $2^{--}$ & & 12971 & 456 & $\chi_{c1}(1P)\Upsilon(1S)$
\\
\cline{3-3}
\cline{5-9}
 & & \multirow{2}{*}{3S} & & $1^{++}$ & \multirow{2}{*}{13566} & 12557 & 1009 & $J/\psi(1S)\Upsilon(1S)$
\\
\cline{5-5}
\cline{7-9}
 & & & & $1^{+-}$ & & 12444 & 1122 & $\eta_{c}(1S)\Upsilon(1S)$
\\
\bottomrule
\end{tabularx}
\end{adjustwidth}
\end{table}

\begin{table}
\caption{Same as in Table~\ref{TabThrCC} but for the $cb\bar c\bar b$ states composed from the scalar (S) diquarks.}
\label{TabThrCB3}
\begin{adjustwidth}{-\extralength}{0cm}
\begin{tabularx}{\fulllength}{CCCCCCCCY{2.5cm}}
\toprule
$\mathbf{QQ\bar Q\bar Q}$ & $\mathbf{d\bar d'}$ & \textbf{State} & $\mathbf{S}$ & $\mathbf{J^{PC}}$ & $\mathbf{M}$ & $\mathbf{M_{thr}}$ & $\boldsymbol\Delta$ & \textbf{Meson pair}
\\
\midrule
\multirow{13}{*}{$cb\bar c\bar b$} & \multirow{13}{*}{$S\bar S$} & 1S & \multirow{13}{*}{0} & $0^{++}$ & 12856 & 12383 & 473 & $\eta_{c}(1S)\eta_{b}(1S)$
\\
\cline{3-3}
\cline{5-9}
 & & \multirow{8}{*}{1P} & & \multirow{8}{*}{\textcolor{black}{$1^{--}$}} & \multirow{8}{*}{\textcolor{black}{13095}} & 12875 & \textcolor{black}{220} & $\chi_{c0}(1P)\Upsilon(1S)$
 \\
\cline{7-9}
 & & & & & & 12883 & 212 & $\eta_{c}(1S)h_{b}(1P)$
 \\
\cline{7-9}
 & & & & & & 12924 & 171 & $h_{c}(1P)\eta_{b}(1S)$
 \\
\cline{7-9}
 & & & & & & 12956 & 139 & $J/\psi(1S)\chi_{b0}(1P)$
 \\
\cline{7-9}
 & & & & & & 12971 & 124 & $\chi_{c1}(1P)\Upsilon(1S)$
 \\
\cline{7-9}
 & & & & & & 12990 & 105 & $J/\psi(1S)\chi_{b1}(1P)$
 \\
\cline{7-9}
 & & & & & & 13009 & 86 & $J/\psi(1S)\chi_{b2}(1P)$
 \\
\cline{7-9}
 & & & & & & 13016 & 79 & $\chi_{c2}(1P)\Upsilon(1S)$
\\
\cline{3-3}
\cline{5-9}
 & & 2S & & $0^{++}$ & 13250 & 12383 & 867 & $\eta_{c}(1S)\eta_{b}(1S)$
\\
\cline{3-3}
\cline{5-9}
 & & 1D & & $2^{++}$ & 13293 & 12557 & 736 & $J/\psi(1S)\Upsilon(1S)$
\\
\cline{3-3}
\cline{5-9}
 & & 2P & & $1^{--}$ & 13420 & 12875 & 545 & $\chi_{c0}(1P)\Upsilon(1S)$
\\
\cline{3-3}
\cline{5-9}
 & & 3S & & $0^{++}$ & 13559 & 12383 & 1176 & $\eta_{c}(1S)\eta_{b}(1S)$
\\
\bottomrule
\end{tabularx}
\end{adjustwidth}
\end{table}

\begin{table} 
\caption{Same as in Table~\ref{TabThrCC} but for the $bb\bar b\bar b$ states composed from the axialvector (A) diquarks.}
\label{TabThrBB}
\begin{adjustwidth}{-\extralength}{0cm}
\begin{tabularx}{\fulllength}{CCCCCCCCY{3.5cm}}
\toprule
$\mathbf{QQ\bar Q\bar Q}$ & $\mathbf{d\bar d'}$ & \textbf{State} & $\mathbf{S}$ & $\mathbf{J^{PC}}$ & $\mathbf{M}$ & $\mathbf{M_{thr}}$ & $\boldsymbol\Delta$ & \textbf{Meson pair}
\\
\midrule
\multirow{46}{*}{$bb\bar b\bar b$} & \multirow{46}{*}{$A\bar A$} & \multirow{4}{*}{1S} & \multirow{2}{*}{0} & \multirow{2}{*}{$\mathbf{0^{++}}$} & \multirow{2}{*}{$\mathbf{19315}$} & 18798 & 517 & $\eta_{b}(1S)\eta_{b}(1S)$
\\
 & & & & & & $\mathbf{18921}$ & $\mathbf{394}$ & \boldmath $\Upsilon(1S)\Upsilon(1S)$
\\
\cline{4-9}
 & & & 1 & $1^{+-}$ & 19320 & 18859 & 461 & $\eta_{b}(1S)\Upsilon(1S)$
\\
\cline{4-9}
 & & & 2 & $\mathbf{2^{++}}$ & $\mathbf{19331}$ & $\mathbf{18921}$ & $\mathbf{410}$ & \boldmath $\Upsilon(1S)\Upsilon(1S)$
\\
\cline{3-9}
 & & \multirow{17}{*}{1P} & \multirow{4}{*}{0} & \multirow{4}{*}{\textcolor{black}{$1^{--}$}} & \multirow{4}{*}{\textcolor{black}{19536}} & 19298 & \textcolor{black}{238} & $\eta_{b}(1S)h_{b}(1P)$
\\
 & & & & & & 19320 & 216 & $\Upsilon(1S)\chi_{b0}(1P)$
\\
 & & & & & & 19353 & 183 & $\Upsilon(1S)\chi_{b1}(1P)$
\\
 & & & & & & 19373 & 163 & $\Upsilon(1S)\chi_{b2}(1P)$
\\
\cline{4-9}
 & & & \multirow{6}{*}{1} & \multirow{2}{*}{\textcolor{black}{$0^{-+}$}} & \multirow{2}{*}{\textcolor{black}{19533}} & 19258 & \textcolor{black}{275} & $\eta_{b}(1S)\chi_{b0}(1P)$
\\
 & & & & & & 19360 & 173 & $\Upsilon(1S)h_{b}(1P)$
\\ 
\cline{5-9}
 & & & & \multirow{2}{*}{\textcolor{black}{$1^{-+}$}} & \multirow{2}{*}{\textcolor{black}{19535}} & 19291 & \textcolor{black}{244} & $\eta_{b}(1S)\chi_{b1}(1P)$
\\
 & & & & & & 19360 & 175 & $\Upsilon(1S)h_{b}(1P)$
\\
\cline{5-9}
 & & & & \multirow{2}{*}{\textcolor{black}{$2^{-+}$}} & \multirow{2}{*}{\textcolor{black}{19539}} & 19311 & \textcolor{black}{228} & $\eta_{b}(1S)\chi_{b2}(1P)$
\\
 & & & & & & 19360 & 179 & $\Upsilon(1S)h_{b}(1P)$
 \\
\cline{4-9}
 & & & \multirow{7}{*}{2} & \multirow{4}{*}{\textcolor{black}{$1^{--}$}} & \multirow{4}{*}{\textcolor{black}{19534}} & 19298 & \textcolor{black}{236} & $\eta_{b}(1S)h_{b}(1P)$
\\
 & & & & & & 19320 & 214 & $\Upsilon(1S)\chi_{b0}(1P)$
\\
 & & & & & & 19353 & 181 & $\Upsilon(1S)\chi_{b1}(1P)$
\\
 & & & & & & 19373 & 161 & $\Upsilon(1S)\chi_{b2}(1P)$
\\
\cline{5-9}
 & & & & \multirow{2}{*}{\textcolor{black}{$2^{--}$}} & \multirow{2}{*}{\textcolor{black}{19538}} & 19353 & \textcolor{black}{185} & $\Upsilon(1S)\chi_{b1}(1P)$
\\
 & & & & & & 19373 & 165 & $\Upsilon(1S)\chi_{b2}(1P)$
\\
\cline{5-9}
 & & & & \textcolor{black}{$3^{--}$} & \textcolor{black}{19545} & 19373 & \textcolor{black}{172} & $\Upsilon(1S)\chi_{b2}(1P)$
\\
\cline{3-9}
 & & \multirow{4}{*}{2S} & \multirow{2}{*}{0} & \multirow{2}{*}{$\mathbf{0^{++}}$} & \multirow{2}{*}{$\mathbf{19680}$} & 18798 & 882 & $\eta_{b}(1S)\eta_{b}(1S)$
\\
 & & & & & & $\mathbf{18921}$ & $\mathbf{759}$ & \boldmath $\Upsilon(1S)\Upsilon(1S)$
\\
\cline{4-9}
 & & & 1 & $1^{+-}$ & 19682 & 18859 & 823 & $\eta_{b}(1S)\Upsilon(1S)$
\\
\cline{4-9}
 & & & 2 & $\mathbf{2^{++}}$ & $\mathbf{19687}$ & $\mathbf{18921}$ & $\mathbf{766}$ & \boldmath $\Upsilon(1S)\Upsilon(1S)$
\\
\cline{3-9}
 &  & \multirow{10}{*}{1D} & 0 & $\mathbf{2^{++}}$ & $\mathbf{19715}$ & $\mathbf{18921}$ & $\mathbf{794}$ & \boldmath $\Upsilon(1S)\Upsilon(1S)$
\\
\cline{4-9}
 & & & \multirow{3}{*}{1} & $1^{+-}$ & 19710 & 18859 & 851 & $\eta_{b}(1S)\Upsilon(1S)$
\\
\cline{5-9}
 & & & & \textcolor{black}{$2^{+-}$} & \textcolor{black}{19714} & 19562 & \textcolor{black}{152} & $\eta_{b}(1S)\Upsilon_{2}(1D)$
\\
\cline{5-9}
 & & & & \textcolor{Maroon}{$3^{+-}$} & \textcolor{Maroon}{19720} & \textcolor{Maroon}{19812} & \textcolor{Maroon}{-92} & \textcolor{Maroon}{$h_{b}(1P)\chi_{b2}(1P)$}
\\
\cline{4-9}
 & & & \multirow{6}{*}{2} & \multirow{2}{*}{$\mathbf{0^{++}}$} & \multirow{2}{*}{$\mathbf{19705}$} & 18798 & 907 & $\eta_{b}(1S)\eta_{b}(1S)$
\\
 & & & & & & $\mathbf{18921}$ & $\mathbf{784}$ & \boldmath $\Upsilon(1S)\Upsilon(1S)$
\\
\cline{5-9}
 & & & & $\mathbf{1^{++}}$ & $\mathbf{19707}$ & $\mathbf{18921}$ & $\mathbf{786}$ & \boldmath $\Upsilon(1S)\Upsilon(1S)$
\\
\cline{5-9}
 & & & & $\mathbf{2^{++}}$ & $\mathbf{19711}$ & $\mathbf{18921}$ & $\mathbf{790}$ & \boldmath $\Upsilon(1S)\Upsilon(1S)$
\\
\cline{5-9}
 & & & & \textcolor{RedViolet}{$3^{++}$} & \textcolor{RedViolet}{19717} & 19624 & \textcolor{RedViolet}{93} & $\Upsilon(1S)\Upsilon_{2}(1D)$
\\
\cline{5-9}
 & & & & \textcolor{Maroon}{$4^{++}$} & \textcolor{Maroon}{19724} & \textcolor{Maroon}{19824} & \textcolor{Maroon}{-100} & \textcolor{Maroon}{$\chi_{b2}(1P)\chi_{b2}(1P)$}
\\
\cline{3-9}
 & & \multirow{7}{*}{2P} & 0 & $1^{--}$ & 19820 & 19298 & 522 & $\eta_{b}(1S)h_{b}(1P)$
\\
\cline{4-9}
 & & & \multirow{3}{*}{1} & $0^{-+}$ & 19821 & 19258 & 563 & $\eta_{b}(1S)\chi_{b0}(1P)$
\\
\cline{5-9}
 & & & & $1^{-+}$ & 19821 & 19291 & 530 & $\eta_{b}(1S)\chi_{b1}(1P)$
\\
\cline{5-9}
 & & & & $2^{-+}$ & 19822 & 19311 & 511 & $\eta_{b}(1S)\chi_{b2}(1P)$
\\
\cline{4-9}
 & & & \multirow{3}{*}{2} & $1^{--}$ & 19823 & 19298 & 525 & $\eta_{b}(1S)h_{b}(1P)$
\\
\cline{5-9}
 & & & & $2^{--}$ & 19823 & 19353 & 470 & $\Upsilon(1S)\chi_{b1}(1P)$
\\
\cline{5-9}
 & & & & $3^{--}$ & 19824 & 19373 & 451 & $\Upsilon(1S)\chi_{b2}(1P)$
\\
\cline{3-9}
 & & \multirow{4}{*}{3S} & \multirow{2}{*}{0} & \multirow{2}{*}{$\mathbf{0^{++}}$} & \multirow{2}{*}{$\mathbf{19941}$} & 18798 & 1143 & $\eta_{b}(1S)\eta_{b}(1S)$
\\
 & & & & & & $\mathbf{18921}$ & $\mathbf{1020}$ & \boldmath $\Upsilon(1S)\Upsilon(1S)$
\\
\cline{4-9}
 & & & 1 & $1^{+-}$ & 19943 & 18859 & 1084 & $\eta_{b}(1S)\Upsilon(1S)$
\\
\cline{4-9}
 & & & 2 & $\mathbf{2^{++}}$ & $\mathbf{19947}$ & $\mathbf{18921}$ & $\mathbf{1026}$ & \boldmath $\Upsilon(1S)\Upsilon(1S)$
\\
\bottomrule
\end{tabularx}
\end{adjustwidth}
\end{table}

\par
From Tables~\ref{TabThrCC}-\ref{TabThrBB} a number of conclusions can be drawn. First of all, with the exception of the two following states:
\begin{align}
\label{EqBB1}
X_{bbbb} && 1D && S=1 && 3^{+-} && 19720 \; MeV
\\
\label{EqBB2}
X_{bbbb} && 1D && S=2 && 4^{++} && 19724 \; MeV, 
\end{align}
for all other tetraquark states there is at least one meson pair with the total mass  less than the tetraquark mass (${ \Delta_{\rm max} > 0}$). Therefore for almost all tetraquarks there is a possibility of such  fall-apart decay.
\par
For most tetraquarks the value of $\Delta_{max}$ significantly exceeds 300 MeV. These tetraquarks lie significantly higher than the decay thresholds and, thus, they rapidly fall-apart in the meson pair due to the quark and antiquark rearrangements. This means that experimentally such a state will manifest itself not as a narrow, but as a wide resonance which is hard to observe. However, such arguments can be applied only to the ground states of tetraquarks. For the excited states there are additional restrictions. In particular, these decays will be suppressed either by the centrifugal barrier between the quark and antiquark (for the orbital excitations), or by the zeros of the wave function (for the radial excitations), or both simultaneously, and therefore such tetraquark states can be observed as narrow resonances.
\par
Next, there are also states for which $\Delta_{max} < 300$ MeV. Such states are close to the meson pair decay threshold and, thus, these fall-apart decays have a small phase. For such states, we show in Tables~\ref{TabThrCC}-\ref{TabThrBB}  not only $\Delta_{max}$, but also all close-lying decay channels and their corresponding $\Delta$ in the range $-50 \leq \Delta \leq 300$ MeV. Small negative $\Delta$ are given because our calculations have a theoretical error.
However, if the value of $\Delta_{\rm max} $ is sufficiently negative, the state cannot decay via the strong fall-apart decay processes into two $Q\bar Q'$ quarkonia, and the main channels will be either a decay due to the heavy quark-antiquark annihilation into gluons with their subsequent hadronization into the lighter hadrons (strongly suppressed according to the Okubo-Zweig-Iizuka rule), or radiative decays (if allowed). As a result, this state will be a narrow state that can be observed experimentally in other decay channels: either to hadrons made up of lighter quarks and antiquarks, or two quarkonia and a photon.
\par
All possible  di-$J/\psi$ and di-$\Upsilon(1S)$ decay thresholds are also given  in  Tables~\ref{TabThrCC}-\ref{TabThrBB} (and highlighted in bold). Such decay channels are the most convenient for the experimental studies, since these mesons have a characteristic decay into a $\mu^{+}\mu^{-}$ pair with branching fractions $ \sim 5 \%$ and, thus, have a clear  experimental signature.
\par
So far, the results of experimental searches are fully correlated with our conclusions. In particular, the LHCb, CMS and ATLAS Collaborations  are searching for the fully-charmed $cc\bar c\bar c$ and fully-bottom $bb\bar b\bar b$ tetraquarks.  In Table~\ref{TabExp} masses and widths of all currently observed fully-charmed tetraquark  states and our candidates for the interpretation of such states are given. One state named X(6900) has already been reliably detected by all three Collaborations (LHCb 2020~\cite{U10-ccLHCb2020}, CMS 2022~\cite{ccCMS2022}, ATLAS 2022~\cite{ccATLAS2022}). It is clearly a candidate for the excited fully-charmed state. Moreover, the measured value of its mass is very close to our prediction. In fact we have 5 candidates for this resonance with the masses within the range of  50 MeV  from the measured X(6900) mass. Thus it is important to measure the quantum numbers of this state (states?). Additionally LHCb data indicates two wide and not very distinctive peaks near 6.4 GeV and 7.2 GeV which can also be interpreted  as ground and excited fully-charmed tetraquark states.
\par
Very recently the CMS~\cite{ccCMS2022} and ATLAS~\cite{ccATLAS2022} Collaborations reported preliminary  results on the observation of exotic charmed states. The CMS Collaboration observed 3 distinct states in the $J/\psi J/\psi$ mass spectrum: X(6600), X(6900) and X(7300), while the ATLAS Collaboration observed 4 distinct states in the di-$J/\psi$ and $J/\psi +\psi(2S)$ channels: X(6200), X(6600), X(6900) and X(7200). As it was already pointed out before, X(6900) is the most prominent of them all since it was observed by all three experiments with very close mass. The peaking structure around 7.2 GeV in LHCb data was confirmed in these experiments (X(7200) and X(7300)). The X(6200) observed by ATLAS is very close to our prediction for the lowest ground state $0^{++}$ with the mass 6190 MeV. The authors of Ref.~\cite{X-6200} also predicted this state from the analysis of the LHCb data back in 2021. For the X(6600) structure observed both by CMS and ATLAS we also propose candidates but with grater deviations from central values of the observed mass.  
\par
On the other hand searches for the fully-bottom tetraquark in the process:
\begin{equation}
\label{EqXbb}
pp \longrightarrow X_{bb\bar b\bar b} \longrightarrow \Upsilon(1S)\Upsilon(1S)
\end{equation}
in the mass range 17.5-20 GeV (covering the mass range we predict: 19.3-20 GeV) have not yet yielded any results (LHCb 2018~\cite{U24LHCBbb-bbNotLHCb2018}, CMS 2017~\cite{bbNotCMS2017}, 2020~\cite{U25CMSbb-bbNotCMS2020}). Moreover, lattice calculations~\cite{U35Bach39bb2018.06} do not find fully-bottom tetraquark bound states in this mass region. Such conclusion correlates with our results that the masses of the fully-bottom tetraquarks are significantly higher than the decay thresholds. Thus these states rapidly fall-apart and can appear only as wide, hard to detect resonances. However, according to our calculations, there are two states of such tetraquarks, corresponding to high orbital excitations with high values of total spin $J$, that lie below any decay thresholds, these are the states already mentioned in~(\ref{EqBB1})-(\ref{EqBB2}). Therefore, these states can be observed as narrow states decaying to lighter hadrons.

\begin{table}[H]
\caption{Exotic X states observed by the LHCb~\cite{U10-ccLHCb2020}, CMS~\cite{ccCMS2022} and ATLAS~\cite{ccATLAS2022} Collaborations in di-$J/\psi$ invariant mass spectra and our candidates. All masses are given in MeV.}
\label{TabExp}
\begin{adjustwidth}{-\extralength}{0cm}
\begin{tabularx}{\fulllength}{>{\rowfont}C>{\rowfont}C>{\rowfont}C>{\rowfont}C>{\rowfont}C>{\rowfont}>{\rowfont}Y{1.0cm}>{\rowfont}C>{\rowfont}C<{\unboldrow}}
\toprule
\boldrow \multirow{2}{*}{Collaboration} & \multirow{2}{*}{State} & \multirow{2}{*}{Mass} & \multirow{2}{*}{Width} & \multicolumn{4}{c}{Our candidates}
\\
\cline{5-8}
\boldrow & & & & State & $\mathbf{S}$ & $\mathbf{J^{PC}}$ & Mass
\\
\midrule
ATLAS & X(6200) & $6220 \pm 50^{+40}_{-50}$ & $310 \pm 120^{+70}_{-80}$ & 1S & 0 & $0^{++}$ & 6190 
\\
\hline
LHCb & X(6400) & $\approx$ 6400 & & 1S & 2 & $2^{++}$ & 6367 
\\
\hline
CMS & \multirow{2}{*}{X(6600)} & $6552 \pm 10 \pm 12$ & $124 \pm 29 \pm 34$ & \multirow{2}{*}{\makecell{1S \\ 2S}} & \multirow{2}{*}{\makecell{2 \\ 0}} & \multirow{2}{*}{\makecell{$2^{++}$ \\ $0^{++}$}} & \multirow{2}{*}{\makecell{6367 \\ 6782}}
\\
\cline{1-1}
\cline{3-4}
ATLAS & & $6620 \pm 30^{+20}_{-10}$ & $310 \pm 90^{+60}_{-110}$ & & & &
\\
\hline
\multirow{2}{*}{LHCb} & \multirow{5}{*}{X(6900)} & $6905 \pm 11 \pm 7$ & $80 \pm 19 \pm 33$ & \multirow{5}{*}{\makecell{2S \\ 1D \\ 1D \\ 1D \\ 1D}} & \multirow{5}{*}{\makecell{2 \\ 0 \\ 2 \\ 2 \\ 2}} & \multirow{5}{*}{\makecell{$2^{++}$ \\ $2^{++}$ \\ $0^{++}$ \\ $1^{++}$ \\ $2^{++}$}} & \multirow{5}{*}{\makecell{6868 \\ 6921 \\ 6899 \\ 6904 \\ 6915}}
\\
\cline{3-4}
 & & $6886 \pm 11 \pm 11$ & $168 \pm 33 \pm 69$ & & & &
\\
\cline{1-1}
\cline{3-4}
\multirow{2}{*}{CMS} & & \multirow{2}{*}{$6927 \pm 9 \pm 5$} & \multirow{2}{*}{$122 \pm 22 \pm 19$} & & & &
\\
& & & & & & & 
\\
\cline{1-1}
\cline{3-4}
ATLAS & & $6870 \pm 30^{+60}_{-10}$ & $120 \pm 40^{+30}_{-10}$ & & & &
\\
\hline
LHCb & \multirow{2}{*}{X(7200)} & $\approx$ 7200 & & \multirow{2}{*}{3S} & \multirow{2}{*}{0} & \multirow{2}{*}{$0^{++}$} & \multirow{2}{*}{7259}
\\
\cline{1-1}
\cline{3-4}
ATLAS & & $7220 \pm 30^{+20}_{-30}$ & $100^{+130+60}_{-70-50}$ & & & &
\\
\hline
CMS & X(7300) & $7287 \pm 19 \pm 5$ & $95 \pm 46 \pm 20$ & \makecell{3S \\ 3S} & \makecell{0 \\ 2} & \makecell{$0^{++}$ \\ $2^{++}$} & \makecell{7259 \\ 7333}
\\
\bottomrule
\end{tabularx}
\end{adjustwidth}
\end{table}

\section{Theoretical predictions}
\label{SecComp}

\par
In Tables~\ref{TabCompcc1S}-\ref{TabCompbb3S} we compare our predictions for the masses (Table~\ref{TabRes}) with the results of other scientific groups obtained in different theoretical approaches.
\par
We have introduced abbreviations in Tables~\ref{TabCompcc1S}-\ref{TabCompbb3S}, but only in cases when the authors used different models or parameter values within the same work. The most common abbreviations are the following.

\begin{itemize}
\item DA, MM, mix -- diquark-antidiquark, meson-meson models and their mixing;

\item (\RNumb{1}-\RNumb{3})$_{d}$ -- different sets of variable parameter values (quark masses, potential parameters, constants, etc.).
\end{itemize}

\noindent
Other abbreviations that occur a few times only:

\begin{itemize}
\item SpB, Osc\RNumb{1},\RNumb{2}~\cite{ExcGroundcc01982} -- Spherical Bag Model and Oscillating Potential Model;

\item QDCSM, ChQM~\cite{U54D40ccbb2020.03} -- Quark Delocalization Color Screening Model and Chiral Quark Model;

\item RSM~\cite{Groundcc1.2020.06} -- Real Scaling Method;

\item Cur, Op~\cite{Groundccbb1.2020.08, Groundccbb3.2021.06} -- different expressions for currents;

\item LO, NLO, NLO$\oplus$G3~\cite{Groundccbb1.2020.08, ExcGroundccbb5.2022.04} -- higher corrections;

\item CQM, CMIM, MCFTM~\cite{U56D38ALL.2021.01} -- Constituent Quark Model, Color-Magnetic Interaction Model and Multiquark Color Flux-Tube Model;

\item K~\cite{GroundALL2.2021.11, GroundALL3.2021.12} -- other geometric configurations of the system that are neither diquark nor meson;

\item NR, Rel~\cite{U44Bach27D30ALL2018.03} -- non-relativistic and relativistic considerations, respectively;

\item Bt, Fl~\cite{Groundccbb6.2022.02} -- ``butterfly'' and ``flip-flop'' potentials.
\end{itemize}

\noindent
A few more clarifying notes to Tables~\ref{TabCompcc1S}-\ref{TabCompbb3S}:

\begin{itemize}

\item In many papers using the diquark-antidiquark picture, the cases of color antitriplet-diquark--triplet-antidiquark $\bar 3 \otimes 3$ and color sextet-diquark--antisextet-antidiquark $6 \otimes \bar 6$  were considered. As we discussed in Sec.~\ref{SecModel}, in the color sextet  (anti)diquark, the interaction potential between (anti)quarks within the (anti)diquark is repulsive and thus corresponding diquark cannot be a bound state, which we consider inappropriate for our problem. Therefore, in Tables~\ref{TabCompcc1S}-\ref{TabCompbb3S} we give theoretical predictions for the masses calculated only for the $\bar 3 \otimes 3$ configuration. If the results were a mixture of two configurations, we chose those masses that contain more of the triplet state in percentage. We note a general trend: in almost all cases, the calculated masses of sextet configurations turned out to be approximately $10-100$ MeV higher than their triplet counterparts.

\item In some papers (for example,~\cite{D34ccbb2020.05}), tetraquarks containing excited diquarks were also considered. Again, as discussed in Sec.~\ref{SecModel}, we have limited ourselves to diquark ground states. Therefore, the masses of such tetraquarks, composed of excited diquarks, are not included in our comparison.

\item *in~\cite{ExcGroundcc01982} for model 1, corrections were calculated only for all 1S states and for the two lowest 1P states; corrections for all  other states were not calculated.

\item **Two cases were considered in~\cite{U51Bach30D39bb2019.02}: the presence and absence of the heavy $\eta_{b}$-meson exchange. The results only for the case without such an exchange are given.

\item ***in~\cite{Groundccbb1.2020.08}, LO results were also obtained, but they are quite similar to NLO$\oplus$G3, so we do not present them.

\item ****in~\cite{ExcGroundccbb5.2022.04}, all results were obtained in two mass schemes: in the $\overline{MS}$-scheme and on-Shell-scheme. In view of the already colossal number of results of this study, we took the masses only in the $\overline{MS}$-scheme.
\end{itemize}



\setlength\LTleft{-\extralength}
\begin{table}[H]
\caption{Same as in Table~\ref{TabCompcc1S} but for the first orbital excitation (1P).}
\label{TabCompcc1P}
\begin{adjustwidth}{-\extralength}{0cm}
\begin{tabularx}{\fulllength}{>{\rowfont}C|>{\rowfont}C>{\rowfont}C>{\rowfont}C>{\rowfont}C>{\rowfont}C>{\rowfont}C>{\rowfont}C<{\unboldrow}}
\toprule
$\mathbf{d\bar d'}$ & \multicolumn{7}{c}{$A\bar A$}
\\
\cline{2-8}
\textbf{State} & \multicolumn{7}{c}{1P}
\\
\cline{2-8}
$\mathbf{S}$ & 0 & \multicolumn{3}{|c|}{1} & \multicolumn{3}{c}{2}
\\
\cline{2-8}
$\mathbf{J^{PC}}$ & $1^{--}$ & \multicolumn{1}{|c}{$0^{-+}$} & \multicolumn{1}{|c}{$1^{-+}$} & \multicolumn{1}{|c|}{$2^{-+}$} & $1^{--}$ & \multicolumn{1}{|c}{$2^{--}$} & \multicolumn{1}{|c}{$3^{--}$}
\\
\midrule
\boldrow Our & 6631 & 6628 & 6634 & 6644 & 6635 & 6648 & 6664
\\
\cite{Exccc01981} & 6550 & 6550 & 6550 & 6550 & 6550 & 6550 & 6550 
\\
\cite{ExcGroundcc01982}*  & \makecell{6694 \scriptsize (Osc\RNumb{1}); \\ 6714 \scriptsize (Osc\RNumb{2})} & \makecell{6695 \scriptsize (Osc\RNumb{1}); \\ 6714 \scriptsize (Osc\RNumb{2})} & \makecell{6718 \scriptsize (Osc\RNumb{1}); \\ 6714 \scriptsize (Osc\RNumb{2})} & \makecell{6718 \scriptsize (Osc\RNumb{1}); \\ 6714 \scriptsize (Osc\RNumb{2})} & \makecell{6718 \scriptsize (Osc\RNumb{1}); \\ 6714 \scriptsize (Osc\RNumb{2})} & \makecell{6718 \scriptsize (Osc\RNumb{1}); \\ 6714 \scriptsize (Osc\RNumb{2})} & \makecell{6718 \scriptsize (Osc\RNumb{1}); \\ 6714 \scriptsize (Osc\RNumb{2})}
\\
\cite{U37Bach19cc2004} & 7004 & 6969 & 7013 & 7033 &  &  &   
\\
\bottomrule
\end{tabularx}
\end{adjustwidth}
\end{table}

\begin{table}[H]
\caption{Same as in Table~\ref{TabCompcc1S} but for the first radial excitation (2S).}
\label{TabCompcc2S}
\begin{adjustwidth}{-\extralength}{0cm}
\begin{tabularx}{\fulllength}{>{\rowfont}Y{3.5cm}|>{\rowfont}C>{\rowfont}C>{\rowfont}C<{\unboldrow}}
\toprule
$\mathbf{d\bar d'}$ & \multicolumn{3}{c}{$A\bar A$}
\\
\cline{2-4}
\textbf{State} & \multicolumn{3}{c}{2S}
\\
\cline{2-4}
$\mathbf{S}$ & 0 & \multicolumn{1}{|c|}{1} & 2
\\
\cline{2-4}
$\mathbf{J^{PC}}$ & $0^{++}$ & \multicolumn{1}{|c|}{$1^{+-}$} & $2^{++}$
\\
\midrule
\boldrow Our & 6782 & 6816 & 6868
\\
\cite{Bach28D31cc2018.04}, \cite{U47Bach31D32cc2019.03} & 6663 & 6675 & 6698
\\
\cite{U48D37ccbb2019.06} &  & \makecell{6856 \scriptsize (\RNumb{1}d); \\ 6894 \scriptsize (\RNumb{2}d)} & \makecell{6864 \scriptsize (\RNumb{1}d); \\ 6919 \scriptsize (\RNumb{2}d)}
\\
\cite{U52Bach33D35ALL2020.01} & 6573 & 6669 & 6739
\\
\bottomrule
\end{tabularx}
\end{adjustwidth}
\end{table}

\begin{table}[H]
\caption{Same as in Table~\ref{TabCompcc1S} but for the second orbital excitation (1D).}
\label{TabCompcc1D}
\begin{adjustwidth}{-\extralength}{0cm}
\begin{tabularx}{\fulllength}{>{\rowfont}C|>{\rowfont}C>{\rowfont}C>{\rowfont}C>{\rowfont}C>{\rowfont}Y{2.3cm}>{\rowfont}C>{\rowfont}C>{\rowfont}C>{\rowfont}C<{\unboldrow}}
\toprule
$\mathbf{d\bar d'}$ & \multicolumn{9}{c}{$A\bar A$}
\\
\cline{2-10}
\textbf{State} & \multicolumn{9}{c}{1D}
\\
\cline{2-10}
$\mathbf{S}$ & 0 & \multicolumn{3}{|c|}{1} & \multicolumn{5}{c}{2}
\\
\cline{2-10}
$\mathbf{J^{PC}}$ & $2^{++}$ & \multicolumn{1}{|c}{$1^{+-}$} & \multicolumn{1}{|c}{$2^{+-}$} & \multicolumn{1}{|c|}{$3^{+-}$} & $0^{++}$ & \multicolumn{1}{|c}{$1^{++}$} & \multicolumn{1}{|c}{$2^{++}$} & \multicolumn{1}{|c}{$3^{++}$} & \multicolumn{1}{|c}{$4^{++}$}
\\
\midrule
\boldrow Our & 6921 & 6909 & 6920 & 6932 & 6899 & 6904 & 6915 & 6929 & 6945
\\
\cite{Exccc01981} & 6780 & 6780 & 6780 & 6780 & 6780 & 6780 & 6780 & 6780 & 6780
\\
\cite{U38Bach20cc2006.01} &  &  & \makecell{6586$\divisionsymbol$6648} &  &  & \makecell{6530$\divisionsymbol$6609} &  &  &  
\\
\cite{U43Bach35D25ccbb2017.03}, \cite{Bach36ccbb2018.05} &  &  &  &  &  & \makecell{$6340 \pm 190$} &  &  &  
\\
\bottomrule
\end{tabularx}
\end{adjustwidth}
\end{table}

\begin{table}[H]
\caption{Same as in Table~\ref{TabCompcc1S} but for the first orbital and radial excitation (2P).}
\label{TabCompcc2P}
\begin{adjustwidth}{-\extralength}{0cm}
\begin{tabularx}{\fulllength}{>{\rowfont}C|>{\rowfont}C>{\rowfont}C>{\rowfont}C>{\rowfont}C>{\rowfont}C>{\rowfont}C>{\rowfont}C<{\unboldrow}}
\toprule
$\mathbf{d\bar d'}$ & \multicolumn{7}{c}{$A\bar A$}
\\
\cline{2-8}
\textbf{State} & \multicolumn{7}{c}{2P}
\\
\cline{2-8}
$\mathbf{S}$ & 0 & \multicolumn{3}{|c|}{1} & \multicolumn{3}{c}{2}
\\
\cline{2-8}
$\mathbf{J^{PC}}$ & $1^{--}$ & \multicolumn{1}{|c}{$0^{-+}$} & \multicolumn{1}{|c}{$1^{-+}$} & \multicolumn{1}{|c|}{$2^{-+}$} & $1^{--}$ & \multicolumn{1}{|c}{$2^{--}$} & \multicolumn{1}{|c}{$3^{--}$}
\\
\midrule
\boldrow Our & 7091 & 7100 & 7099 & 7098 & 7113 & 7113 & 7112
\\
\cite{Bach28D31cc2018.04}, \cite{U47Bach31D32cc2019.03} & 6944 & 6867 & 6944 & 6970 & 6876 & 6962 & 6997
\\
\cite{U52Bach33D35ALL2020.01} & 6940 & 6953 &  &  & 6943 &  &  
\\
\cite{Exccc1.2020.12} & \makecell{$6580 \pm 90$} &  &  &  &  &  &  
\\
\cite{ExcGroundcc3.2021.13} & 6925 & 6851 & 6926 & 6951 & 6849 & 6944 & 6982
\\
\cite{ExcGroundcc6.2021.16} & \makecell{7143 \scriptsize (\RNumb{1}d); \\ 8174 \scriptsize (\RNumb{2}d)} & \makecell{7130 \scriptsize (\RNumb{1}d); \\ 8162 \scriptsize (\RNumb{2}d)} & \makecell{7130 \scriptsize (\RNumb{1}d); \\ 8162 \scriptsize (\RNumb{2}d)} & \makecell{7130 \scriptsize (\RNumb{1}d); \\ 8162 \scriptsize (\RNumb{2}d)} & \makecell{7134 \scriptsize (\RNumb{1}d); \\ 8166 \scriptsize (\RNumb{2}d)} & \makecell{7134 \scriptsize (\RNumb{1}d); \\ 8166 \scriptsize (\RNumb{2}d)} & \makecell{7134 \scriptsize (\RNumb{1}d); \\ 8166 \scriptsize (\RNumb{2}d)}
\\
\cite{ExcGroundcc7.2022.05} & $6740 \pm 90$ &  &  &  &  &  &  
\\
\bottomrule
\end{tabularx}
\end{adjustwidth}
\end{table}

\begin{table}[H]
\caption{Same as in Table~\ref{TabCompcc1S} but for the second radial excitation (3S).}
\label{TabCompcc3S}
\begin{adjustwidth}{-\extralength}{0cm}
\begin{tabularx}{\fulllength}{>{\rowfont}C|>{\rowfont}C>{\rowfont}C>{\rowfont}C<{\unboldrow}}
\toprule
$\mathbf{d\bar d'}$ & \multicolumn{3}{c}{$A\bar A$}
\\
\cline{2-4}
\textbf{State} & \multicolumn{3}{c}{3S}
\\
\cline{2-4}
$\mathbf{S}$ & 0 & \multicolumn{1}{|c|}{1} & 2
\\
\cline{2-4}
$\mathbf{J^{PC}}$ & $0^{++}$ & \multicolumn{1}{|c|}{$1^{+-}$} & $2^{++}$
\\
\midrule
\boldrow Our & 7259 & 7287 & 7333
\\
\cite{U48D37ccbb2019.06} &  & \makecell{6915 \scriptsize (\RNumb{1}d); \\ 7036 \scriptsize (\RNumb{2}d)} & \makecell{6919 \scriptsize (\RNumb{1}d); \\ 7058 \scriptsize (\RNumb{2}d)}
\\
\cite{U52Bach33D35ALL2020.01} & 6948 & 7016 & 7071
\\
\cite{U54D40ccbb2020.03} & \makecell{7225, 7250 \scriptsize (DA, QDCSM (\RNumb{1}-\RNumb{3})$_{d}$); \\ 7210, 7250, 7260 \scriptsize (DA, ChQM (\RNumb{1}-\RNumb{3})$_{d}$)} &  &  
\\
\bottomrule
\end{tabularx}
\end{adjustwidth}
\end{table}

\begin{table}[H]
\caption{Same as in Table~\ref{TabCompcc1S} but for the ground state (1S) of $cb\bar c\bar b$ composed from the axialvector diquark and antidiquark.}
\label{TabCompcb1S}
\begin{adjustwidth}{-\extralength}{0cm}
\begin{tabularx}{\fulllength}{>{\rowfont}C|>{\rowfont}C>{\rowfont}C>{\rowfont}C<{\unboldrow}}
\toprule
$\mathbf{d\bar d'}$ & \multicolumn{3}{c}{$A\bar A$}
\\
\cline{2-4}
\textbf{State} & \multicolumn{3}{c}{1S}
\\
\cline{2-4}
$\mathbf{S}$ & 0 & \multicolumn{1}{|c|}{1} & 2
\\
\cline{2-4}
$\mathbf{J^{PC}}$ & $0^{++}$ & \multicolumn{1}{|c|}{$1^{+-}$} & $2^{++}$
\\
\midrule
\boldrow Our & 12838 & 12855 & 12883
\\
\cite{U41Bach22D23ccbbcb2012.01} & 12471 & 12488 & 12566 
\\
\cite{U46Bach26D29ALL2018.02} & 13483 & 13592 & 13590 
\\
\cite{U44Bach27D30ALL2018.03} & $\lesssim 12620$ &  &   
\\
\bottomrule
\end{tabularx}
\end{adjustwidth}
\end{table}

\begin{table}[H]
\caption{Same as in Table~\ref{TabCompcc1S} but for the first orbital excitation (1P) of $cb\bar c\bar b$ composed from the axialvector diquark and antidiquark.}
\label{TabCompcb1P}
\begin{adjustwidth}{-\extralength}{0cm}
\begin{tabularx}{\fulllength}{>{\rowfont}C|>{\rowfont}C>{\rowfont}C>{\rowfont}C>{\rowfont}C>{\rowfont}C>{\rowfont}C>{\rowfont}C<{\unboldrow}}
\toprule
$\mathbf{d\bar d'}$ & \multicolumn{7}{c}{$A\bar A$}
\\
\cline{2-8}
\textbf{State} & \multicolumn{7}{c}{1P}
\\
\cline{2-8}
$\mathbf{S}$ & 0 & \multicolumn{3}{|c|}{1} & \multicolumn{3}{c}{2}
\\
\cline{2-8}
$\mathbf{J^{PC}}$ & $1^{--}$ & \multicolumn{1}{|c}{$0^{-+}$} & \multicolumn{1}{|c}{$1^{-+}$} & \multicolumn{1}{|c|}{$2^{-+}$} & $1^{--}$ & \multicolumn{1}{|c}{$2^{--}$} & \multicolumn{1}{|c}{$3^{--}$}
\\
\midrule
\boldrow Our & 13103 & 13100 & 13103 & 13108 & 13103 & 13109 & 13116
\\
\cite{U52Bach33D35ALL2020.01} & 12934 & 12943 &  &  & 12944 &  &  
\\
\bottomrule
\end{tabularx}
\end{adjustwidth}
\end{table}

\begin{table}[H]
\caption{Same as in Table~\ref{TabCompcc1S} but for the first radial excitation (2S) of $cb\bar c\bar b$ composed from the axialvector diquark and antidiquark.}
\label{TabCompcb2S}
\begin{adjustwidth}{-\extralength}{0cm}
\begin{tabularx}{\fulllength}{>{\rowfont}C|>{\rowfont}C>{\rowfont}C>{\rowfont}C<{\unboldrow}}
\toprule
$\mathbf{d\bar d'}$ & \multicolumn{3}{c}{$A\bar A$}
\\
\cline{2-4}
\textbf{State} & \multicolumn{3}{c}{2S}
\\
\cline{2-4}
$\mathbf{S}$ & 0 & \multicolumn{1}{|c|}{1} & 2
\\
\cline{2-4}
$\mathbf{J^{PC}}$ & $0^{++}$ & \multicolumn{1}{|c|}{$1^{+-}$} & $2^{++}$
\\
\midrule
\boldrow Our & 13247 & 13256 & 13272 
\\
\cite{U52Bach33D35ALL2020.01} & 12975 & 13022 & 13063 
\\
\rule{0pt}{12pt}\cite{ExcGroundccbbcb1.2021.23} & $12911^{+48}_{-51}$ &  & $12914^{+49}_{-49}$  
\\
\bottomrule
\end{tabularx}
\end{adjustwidth}
\end{table}

\begin{table}[H]
\caption{Same as in Table~\ref{TabCompcc1S} but for the second orbital excitation (1D) of $cb\bar c\bar b$ composed from the axialvector diquark and antidiquark.}
\label{TabCompcb1D}
\begin{adjustwidth}{-\extralength}{0cm}
\begin{tabularx}{\fulllength}{>{\rowfont}C|>{\rowfont}C>{\rowfont}C>{\rowfont}C>{\rowfont}C>{\rowfont}C>{\rowfont}C>{\rowfont}C>{\rowfont}C>{\rowfont}C<{\unboldrow}}
\toprule
$\mathbf{d\bar d'}$ & \multicolumn{9}{c}{$A\bar A$}
\\
\cline{2-10}
\textbf{State} & \multicolumn{9}{c}{1D}
\\
\cline{2-10}
$\mathbf{S}$ & 0 & \multicolumn{3}{|c|}{1} & \multicolumn{5}{c}{2}
\\
\cline{2-10}
$\mathbf{J^{PC}}$ & $2^{++}$ & \multicolumn{1}{|c}{$1^{+-}$} & \multicolumn{1}{|c}{$2^{+-}$} & \multicolumn{1}{|c|}{$3^{+-}$} & $0^{++}$ & \multicolumn{1}{|c}{$1^{++}$} & \multicolumn{1}{|c}{$2^{++}$} & \multicolumn{1}{|c}{$3^{++}$} & \multicolumn{1}{|c}{$4^{++}$}
\\
\midrule
\boldrow Our & 13306 & 13299 & 13304 & 13311 & 13293 & 13296 & 13301 & 13308 & 13317
\\
\cite{U52Bach33D35ALL2020.01} & 13166 & 13167 &  &  & 13170 & 13168 & 13166 &  &   
\\
\bottomrule
\end{tabularx}
\end{adjustwidth}
\end{table}

\begin{table}[H]
\caption{Same as in Table~\ref{TabCompcc1S} but for the first orbital and radial excitation (2P) of $cb\bar c\bar b$ composed from the axialvector diquark and antidiquark.}
\label{TabCompcb2P}
\begin{adjustwidth}{-\extralength}{0cm}
\begin{tabularx}{\fulllength}{>{\rowfont}C|>{\rowfont}C>{\rowfont}C>{\rowfont}C>{\rowfont}C>{\rowfont}C>{\rowfont}C>{\rowfont}C<{\unboldrow}}
\toprule
$\mathbf{d\bar d'}$ & \multicolumn{7}{c}{$A\bar A$}
\\
\cline{2-8}
\textbf{State} & \multicolumn{7}{c}{2P}
\\
\cline{2-8}
$\mathbf{S}$ & 0 & \multicolumn{3}{|c|}{1} & \multicolumn{3}{c}{2}
\\
\cline{2-8}
$\mathbf{J^{PC}}$ & $1^{--}$ & \multicolumn{1}{|c}{$0^{-+}$} & \multicolumn{1}{|c}{$1^{-+}$} & \multicolumn{1}{|c|}{$2^{-+}$} & $1^{--}$ & \multicolumn{1}{|c}{$2^{--}$} & \multicolumn{1}{|c}{$3^{--}$}
\\
\midrule
\boldrow Our & 13428 & 13431 & 13431 & 13431 & 13434 & 13435 & 13436 
\\
\cite{U52Bach33D35ALL2020.01} & 13262 & 13269 &  &  & 13269 &  &    
\\
\bottomrule
\end{tabularx}
\end{adjustwidth}
\end{table}

\begin{table}[H]
\caption{Same as in Table~\ref{TabCompcc1S} but for the second radial excitation (3S) of $cb\bar c\bar b$ composed from the axialvector diquark and antidiquark.}
\label{TabCompcb3S}
\begin{adjustwidth}{-\extralength}{0cm}
\begin{tabularx}{\fulllength}{>{\rowfont}C|>{\rowfont}C>{\rowfont}C>{\rowfont}C<{\unboldrow}}
\toprule
$\mathbf{d\bar d'}$ & \multicolumn{3}{c}{$A\bar A$}
\\
\cline{2-4}
\textbf{State} & \multicolumn{3}{c}{3S}
\\
\cline{2-4}
$\mathbf{S}$ & 0 & \multicolumn{1}{|c|}{1} & 2
\\
\cline{2-4}
$\mathbf{J^{PC}}$ & $0^{++}$ & \multicolumn{1}{|c|}{$1^{+-}$} & $2^{++}$
\\
\midrule
\boldrow Our & 13558 & 13566 & 13580
\\
\cite{U52Bach33D35ALL2020.01} & 13301 & 13335 & 13365
\\
\rule{0pt}{12pt}\cite{ExcGroundccbbcb1.2021.23} & $13200^{+35}_{-36}$ &  & $13202^{+35}_{-36}$
\\
\bottomrule
\end{tabularx}
\end{adjustwidth}
\end{table}

\begin{table}[H]
\caption{Same as in Table~\ref{TabCompcc1S} but for the ground state (1S) first orbital excitation (1P) and first radial excitation (2S) of $cb\bar c\bar b$ composed from the mixture of axialvector and scalar diquark and antidiquark.}
\label{TabCompcbAS1}
\begin{adjustwidth}{-\extralength}{0cm}
\begin{tabularx}{\fulllength}{>{\rowfont}C|>{\rowfont}C>{\rowfont}C>{\rowfont}C>{\rowfont}C>{\rowfont}C<{\unboldrow}}
\toprule
$\mathbf{d\bar d'}$ & \multicolumn{5}{c}{$\dfrac{1}{\sqrt{2}} \Big( A\bar S \pm S\bar A \Big)$}
\\
\cline{2-6}
\textbf{State} & 1S & \multicolumn{3}{|c|}{1P} & 2S
\\
\cline{2-6}
$\mathbf{S}$ & \multicolumn{5}{c}{1}
\\
\cline{2-6}
$\mathbf{J^{PC}}$ & $1^{+\pm}$ & \multicolumn{1}{|c}{$0^{-\pm}$} & \multicolumn{1}{|c}{$1^{-\pm}$} & \multicolumn{1}{|c}{$2^{-\pm}$} & \multicolumn{1}{|c}{$1^{+\pm}$}
\\
\midrule
\boldrow Our & 12863 & 13096 & 13099 & 13104 & 13257
\\
\cite{U41Bach22D23ccbbcb2012.01} & \makecell{12485 (+); \\ 12424 (-)} &  &  &  &  
\\
\cite{U46Bach26D29ALL2018.02} & \makecell{13599 (+); \\ 13555 (-)} &  &  &  &  
\\
\cite{U49Bach29D33ALL2019.01} & \makecell{12870 (+); \\ 12852 (-)} &  &  &  &  
\\
\bottomrule
\end{tabularx}
\end{adjustwidth}
\end{table}

\begin{table}[H]
\caption{Same as in Table~\ref{TabCompcc1S} but for the second orbital excitation (1D) first orbital and radial excitation (2P) and second radial excitation(3S) of $cb\bar c\bar b$ composed from the mixture of axialvector and scalar diquark and antidiquark.}
\label{TabCompcbAS2}
\begin{adjustwidth}{-\extralength}{0cm}
\begin{tabularx}{\fulllength}{>{\rowfont}C|>{\rowfont}C>{\rowfont}C>{\rowfont}C>{\rowfont}C>{\rowfont}C>{\rowfont}C>{\rowfont}C<{\unboldrow}}
\toprule
$\mathbf{d\bar d'}$ & \multicolumn{7}{c}{$\dfrac{1}{\sqrt{2}} \Big( A\bar S \pm S\bar A \Big)$}
\\
\cline{2-8}
\textbf{State} & \multicolumn{3}{c}{1D} & \multicolumn{3}{|c|}{2P} & 3S
\\
\cline{2-8}
$\mathbf{S}$ & \multicolumn{7}{c}{1}
\\
\cline{2-8}
$\mathbf{J^{PC}}$ & $1^{+\pm}$ & \multicolumn{1}{|c}{$2^{+\pm}$} & \multicolumn{1}{|c|}{$3^{+\pm}$} & $0^{-\pm}$ & \multicolumn{1}{|c}{$1^{-\pm}$} & \multicolumn{1}{|c}{$2^{-\pm}$} & \multicolumn{1}{|c}{$1^{+\pm}$}
\\
\midrule
\boldrow Our & 13293 & 13298 & 13305 & 13426 & 13426 & 13427 & 13566
\\
\cite{U52Bach33D35ALL2020.01} & 13154 &  &  & 13250 & 13250 (-) &  & 13342
\\
\bottomrule
\end{tabularx}
\end{adjustwidth}
\end{table}

\begin{table}[H]
\caption{Same as in Table~\ref{TabCompcc1S} but for the ground state (1S), first orbital excitation (1P), first radial excitation (2S), second orbital excitation (1D), first orbital and radial excitation (2P), second radial excitation (3S) of $cb\bar c\bar b$ composed from the scalar diquark and antidiquark.}
\label{TabCompcbSS}
\begin{adjustwidth}{-\extralength}{0cm}
%

\begin{table}[H]
\caption{Same as in Table~\ref{TabCompcc1S} but for the first orbital excitation (1P) of the $bb\bar b\bar b$.}
\label{TabCompbb1P}
\begin{adjustwidth}{-\extralength}{0cm}
\begin{tabularx}{\fulllength}{>{\rowfont}C|>{\rowfont}C>{\rowfont}C>{\rowfont}C>{\rowfont}C>{\rowfont}C>{\rowfont}C>{\rowfont}C<{\unboldrow}}
\toprule
$\mathbf{d\bar d'}$ & \multicolumn{7}{c}{$A\bar A$}
\\
\cline{2-8}
\textbf{State} & \multicolumn{7}{c}{1P}
\\
\cline{2-8}
$\mathbf{S}$ & 0 & \multicolumn{3}{|c|}{1} & \multicolumn{3}{c}{2}
\\
\cline{2-8}
$\mathbf{J^{PC}}$ & $1^{--}$ & \multicolumn{1}{|c}{$0^{-+}$} & \multicolumn{1}{|c}{$1^{-+}$} & \multicolumn{1}{|c|}{$2^{-+}$} & $1^{--}$ & \multicolumn{1}{|c}{$2^{--}$} & \multicolumn{1}{|c}{$3^{--}$}
\\
\midrule
\boldrow Our & 19536 & 19533 & 19535 & 19539 & 19534 & 19538 & 19545
\\
\cite{U43Bach35D25ccbb2017.03}, \cite{Bach36ccbb2018.05} & \makecell{$18770 \pm 160$} & \makecell{$18790 \pm 180$} & \makecell{$18830 \pm 180$} &  &  &  & 
\\
\cite{Bach37D28ccbb2019.05} & \makecell{$18890 \pm 90$} &  &  &  &  &  &  
\\
\cite{U52Bach33D35ALL2020.01} & 19281 & 19288 &  &  & 19288 &  &  
\\
\bottomrule
\end{tabularx}
\end{adjustwidth}
\end{table}

\begin{table}[H]
\caption{Same as in Table~\ref{TabCompcc1S} but for the first radial excitation (2S) of the $bb\bar b\bar b$.}
\label{TabCompbb2S}
\begin{adjustwidth}{-\extralength}{0cm}
\begin{tabularx}{\fulllength}{>{\rowfont}C|>{\rowfont}C>{\rowfont}C>{\rowfont}C<{\unboldrow}}
\toprule
$\mathbf{d\bar d'}$ & \multicolumn{3}{c}{$A\bar A$}
\\
\cline{2-4}
\textbf{State} & \multicolumn{3}{c}{2S}
\\
\cline{2-4}
$\mathbf{S}$ & 0 & \multicolumn{1}{|c|}{1} & 2
\\
\cline{2-4}
$\mathbf{J^{PC}}$ & $0^{++}$ & \multicolumn{1}{|c|}{$1^{+-}$} & $2^{++}$
\\
\midrule
\boldrow Our & 19680 & 19682 & 19687
\\
\cite{U48D37ccbb2019.06} &  & \makecell{19594 \scriptsize (\RNumb{1}d); \\ 19813 \scriptsize  (\RNumb{2}d)} & \makecell{19596 \scriptsize (\RNumb{1}d); \\ 19823 \scriptsize (\RNumb{2}d)}
\\
\cite{U52Bach33D35ALL2020.01} & 19335 & 19366 & 19398
\\
\cite{U53D9ccbb2020.02} & $19434 \pm 25$ &  & $19481 \pm 25$
\\
\bottomrule
\end{tabularx}
\end{adjustwidth}
\end{table}

\begin{table}[H]
\caption{Same as in Table~\ref{TabCompcc1S} but for the second orbital excitation (1D) of the $bb\bar b\bar b$.}
\label{TabCompbb1D}
\begin{adjustwidth}{-\extralength}{0cm}
\begin{tabularx}{\fulllength}{>{\rowfont}C|>{\rowfont}C>{\rowfont}C>{\rowfont}C>{\rowfont}C>{\rowfont}C>{\rowfont}C>{\rowfont}C>{\rowfont}C>{\rowfont}C<{\unboldrow}}
\toprule
$\mathbf{d\bar d'}$ & \multicolumn{9}{c}{$A\bar A$}
\\
\cline{2-10}
\textbf{State} & \multicolumn{9}{c}{1D}
\\
\cline{2-10}
$\mathbf{S}$ & 0 & \multicolumn{3}{|c|}{1} & \multicolumn{5}{c}{2}
\\
\cline{2-10}
$\mathbf{J^{PC}}$ & $2^{++}$ & \multicolumn{1}{|c}{$1^{+-}$} & \multicolumn{1}{|c}{$2^{+-}$} & \multicolumn{1}{|c|}{$3^{+-}$} & $0^{++}$ & \multicolumn{1}{|c}{$1^{++}$} & \multicolumn{1}{|c}{$2^{++}$} & \multicolumn{1}{|c}{$3^{++}$} & \multicolumn{1}{|c}{$4^{++}$}
\\
\midrule
\boldrow Our & 19715 & 19710 & 19714 & 19720 & 19705 & 19707 & 19711 & 19717 & 19724
\\
\cite{U43Bach35D25ccbb2017.03}, \cite{Bach36ccbb2018.05} &  &  &  &  &  & \makecell{$18320 \pm  180$} &  &  &  
\\
\cite{U52Bach33D35ALL2020.01} & 19510 & 19511 &  &  & 19513 & 19512 & 19510 &  &  
\\
\cite{Excccbb1.2021.24} & 19669 & 19671 & 19672 & 19675 & 19677 & 19678 & 19680 & 19684 & 19686
\\
\rule{0pt}{12pt} \cite{ExcGroundccbb5.2022.04}****  &  &  &  &  &  & \makecell{$19200^{+200}_{-260}$ \\ \scriptsize (DA, LO); \\ $18850^{+190}_{-260}$ \\ \scriptsize (DA, NLO); \\ $18500^{+170}_{-260}$\\$\divisionsymbol 19210^{+200}_{-260}$ \\ \scriptsize (MM, LO); \\ $18600^{+190}_{-260}$\\$\divisionsymbol 19530^{+110}_{-170}$ \\ \scriptsize (MM, NLO)} &  &  &  
\\
\bottomrule
\end{tabularx}
\end{adjustwidth}
\end{table}

\begin{table}[H]
\caption{Same as in Table~\ref{TabCompcc1S} but for the second orbital and radial excitation (2P) of the $bb\bar b\bar b$.}
\label{TabCompbb2P}
\begin{adjustwidth}{-\extralength}{0cm}
\begin{tabularx}{\fulllength}{>{\rowfont}C|>{\rowfont}C>{\rowfont}C>{\rowfont}C>{\rowfont}C>{\rowfont}C>{\rowfont}C>{\rowfont}C<{\unboldrow}}
\toprule
$\mathbf{d\bar d'}$ & \multicolumn{7}{c}{$A\bar A$}
\\
\cline{2-8}
\textbf{State} & \multicolumn{7}{c}{2P}
\\
\cline{2-8}
$\mathbf{S}$ & 0 & \multicolumn{3}{|c|}{1} & \multicolumn{3}{c}{2}
\\
\cline{2-8}
$\mathbf{J^{PC}}$ & $1^{--}$ & \multicolumn{1}{|c}{$0^{-+}$} & \multicolumn{1}{|c}{$1^{-+}$} & \multicolumn{1}{|c|}{$2^{-+}$} & $1^{--}$ & \multicolumn{1}{|c}{$2^{--}$} & \multicolumn{1}{|c}{$3^{--}$}
\\
\midrule
\boldrow Our & 19820 & 19821 & 19821 & 19822 & 19823 & 19823 & 19824 
\\
\cite{U52Bach33D35ALL2020.01} & 19597 & 19602 &  &  & 19602 &  &   
\\
\bottomrule
\end{tabularx}
\end{adjustwidth}
\end{table}

\begin{table}[H]
\caption{Same as in Table~\ref{TabCompcc1S} but for the third radial excitation (3S) of the $bb\bar b\bar b$.}
\label{TabCompbb3S}
\begin{adjustwidth}{-\extralength}{0cm}
\begin{tabularx}{\fulllength}{>{\rowfont}C|>{\rowfont}C>{\rowfont}C>{\rowfont}C<{\unboldrow}}
\toprule
$\mathbf{d\bar d'}$ & \multicolumn{3}{c}{$A\bar A$}
\\
\cline{2-4}
\textbf{State} & \multicolumn{3}{c}{3S}
\\
\cline{2-4}
$\mathbf{S}$ & 0 & \multicolumn{1}{|c|}{1} & 2
\\
\cline{2-4}
$\mathbf{J^{PC}}$ & $0^{++}$ & \multicolumn{1}{|c|}{$1^{+-}$} & $2^{++}$
\\
\midrule
\boldrow Our & 19941 & 19943 & 19947
\\
\cite{U48D37ccbb2019.06} &  & \makecell{19681 \scriptsize (\RNumb{1}d); \\ 20065 \scriptsize (\RNumb{2}d)} & \makecell{19682 \scriptsize (\RNumb{1}d); \\ 20077 \scriptsize (\RNumb{2d})} 
\\
\cite{U52Bach33D35ALL2020.01} & 19644 & 19665 & 19688
\\
\cite{U55D41ccbb2020.04} & 19726 & 19733 & 19736
\\
\bottomrule
\end{tabularx}
\end{adjustwidth}
\end{table}

\par
We compare our predictions with the results obtained in the following approaches and models:

\begin{itemize}
\item Various quark models:~\cite{Groundcc01975, Exccc01981, ExcGroundcc01982, U37Bach19cc2004, U38Bach20cc2006.01, U39Bach21cc2011, U41Bach22D23ccbbcb2012.01, U40Bach23cc2012.02, Bach24D26ccbb2017.01, U45Bach25bb2018.01, U46Bach26D29ALL2018.02, U44Bach27D30ALL2018.03, Bach28D31cc2018.04, U49Bach29D33ALL2019.01, U51Bach30D39bb2019.02, U47Bach31D32cc2019.03, U50Bach32D36bb2019.04, U48D37ccbb2019.06, D42cb2019.07, U52Bach33D35ALL2020.01, U53D9ccbb2020.02, U54D40ccbb2020.03, U55D41ccbb2020.04, D34ccbb2020.05, Groundcc1.2020.06, Groundccbb0.2020.07, Groundccbb2.2020.09, GroundALL0.2020.10, ExcGroundccbb1.2020.11, U56D38ALL.2021.01, Groundcc3.2021.03,  Groundbb1.2021.05, Groundccbb4.2021.07, Groundccbb5.2021.08, GroundALL1.2021.10, GroundALL2.2021.11, GroundALL3.2021.12, ExcGroundcc3.2021.13, ExcGroundcc4.2021.14, ExcGroundcc5.2021.15, ExcGroundcc6.2021.16, ExcGroundbb1.2021.18, ExcGroundccbb2.2021.20, ExcGroundccbb4.2021.22, ExcGroundccbbcb1.2021.23, Excccbb1.2021.24, Groundcc5.2022.01, Groundccbb6.2022.02, Groundccbbcb4.2022.03, Groundcc6.2022.06, GroundALL4.2022.07, GroundALL5.2022.08}.

\item QCD sum rules:~\cite{U42Bach34D27ccbb2017.02, U43Bach35D25ccbb2017.03, Bach36ccbb2018.05, Bach37D28ccbb2019.05, Groundccbb1.2020.08, Exccc1.2020.12, Groundcc2.2021.02, Groundcc4.2021.04, Groundccbb3.2021.06, Groundccbbcb3.2021.09, ExcGroundcb1.2021.19, ExcGroundccbb3.2021.21, ExcGroundccbb5.2022.04, ExcGroundcc7.2022.05}.

\item Lattice calculations:~\cite{Exccc0.5.2006.02, U35Bach39bb2018.06}.
\end{itemize}

\noindent
Among them the following configurations can be distinguished: 

\begin{itemize}
\item Diquark-antidiquark model:~\cite{Exccc01981, ExcGroundcc01982, Exccc0.5.2006.02, U39Bach21cc2011, U41Bach22D23ccbbcb2012.01, Bach24D26ccbb2017.01, U42Bach34D27ccbb2017.02, U43Bach35D25ccbb2017.03, U45Bach25bb2018.01, U46Bach26D29ALL2018.02, U44Bach27D30ALL2018.03, Bach28D31cc2018.04, Bach36ccbb2018.05, U35Bach39bb2018.06, U49Bach29D33ALL2019.01, U51Bach30D39bb2019.02, U47Bach31D32cc2019.03, U50Bach32D36bb2019.04, Bach37D28ccbb2019.05, U48D37ccbb2019.06, D42cb2019.07, U52Bach33D35ALL2020.01, U53D9ccbb2020.02, U54D40ccbb2020.03, U55D41ccbb2020.04, D34ccbb2020.05, Groundcc1.2020.06, Groundccbb0.2020.07, Groundccbb1.2020.08, Groundccbb2.2020.09, GroundALL0.2020.10, U56D38ALL.2021.01, Exccc1.2020.12, Groundcc2.2021.02, Groundcc4.2021.04, Groundccbb4.2021.07, Groundccbb5.2021.08, GroundALL1.2021.10, GroundALL2.2021.11, GroundALL3.2021.12, ExcGroundcc3.2021.13, ExcGroundcc4.2021.14, ExcGroundcc5.2021.15, ExcGroundbb1.2021.18, ExcGroundcb1.2021.19, ExcGroundccbb2.2021.20, ExcGroundccbb4.2021.22, ExcGroundccbbcb1.2021.23, Excccbb1.2021.24, Groundccbb6.2022.02, Groundccbbcb4.2022.03, ExcGroundccbb5.2022.04, ExcGroundcc7.2022.05, Groundcc6.2022.06, GroundALL4.2022.07, GroundALL5.2022.08}.

\item Meson-meson model:~\cite{ExcGroundcc01982, Exccc0.5.2006.02, U51Bach30D39bb2019.02, D42cb2019.07, U54D40ccbb2020.03, Groundcc1.2020.06, Groundccbb1.2020.08, Groundccbb2.2020.09, Groundccbbcb3.2021.09, GroundALL2.2021.11, GroundALL3.2021.12, ExcGroundcc6.2021.16, ExcGroundccbb3.2021.21, ExcGroundccbb5.2022.04, Groundcc6.2022.06}.

\item Mixing of the diquark and meson structures:~\cite{U40Bach23cc2012.02, U51Bach30D39bb2019.02, D42cb2019.07, Groundcc1.2020.06, Groundccbb2.2020.09, ExcGroundccbb1.2020.11}.
\end{itemize}

\par
It is seen  from the Tables~\ref{TabCompcc1S}-\ref{TabCompbb3S} that our results agree well (within the $\pm 50$ MeV range) with the following results:

\begin{itemize}
\item For the $cc\bar c\bar c$ tetraquark:

\begin{itemize}
\item in the diquark-antidiquark picture:~\cite{Bach24D26ccbb2017.01, U44Bach27D30ALL2018.03, Groundccbb0.2020.07, Groundcc3.2021.03, Groundccbb5.2021.08} (all  predictions);~\cite{U53D9ccbb2020.02, ExcGroundccbb4.2021.22, ExcGroundcc7.2022.05} (ground states only);~\cite{U43Bach35D25ccbb2017.03, Bach36ccbb2018.05, U52Bach33D35ALL2020.01, D34ccbb2020.05} (1P);~\cite{U48D37ccbb2019.06} (2S);~\cite{ExcGroundcc5.2021.15, Excccbb1.2021.24} (3S);~\cite{U54D40ccbb2020.03} (1S, 3S);~\cite{Groundccbb4.2021.07} (2S, 3S);~\cite{ExcGroundccbb5.2022.04} (1P, 1D).

\item in the other models: \cite{Groundcc01975, U54D40ccbb2020.03, Groundccbb1.2020.08, Groundcc3.2021.03, Groundccbb3.2021.06, GroundALL2.2021.11, GroundALL3.2021.12, ExcGroundccbb5.2022.04} (all  predictions);~\cite{ExcGroundcc01982, ExcGroundccbb3.2021.21} (ground states only);~\cite{ExcGroundcc6.2021.16} (2P);~\cite{ExcGroundccbb1.2020.11} (3S).
\end{itemize}

\item For the $cb\bar c\bar b$ tetraquark:
\begin{itemize}

\item in the diquark-antidiquark picture: $A\bar A, \; \frac{1}{\sqrt{2}}(A\bar S \pm S\bar A), \; S\bar S \;$ --~\cite{U49Bach29D33ALL2019.01, GroundALL2.2021.11, GroundALL3.2021.12, GroundALL4.2022.07} (all  predictions).

\item in other models: $A\bar A$ --~\cite{Groundccbbcb3.2021.09, GroundALL2.2021.11, GroundALL3.2021.12} (all predictions);
$\frac{1}{\sqrt{2}}(A\bar S \pm S\bar A), \; S\bar S \;$ --~\cite{GroundALL2.2021.11, GroundALL3.2021.12} (all  predictions).
\end{itemize}

\item For the $bb\bar b\bar b$ tetraquark:
\begin{itemize}

\item in the diquark-antidiquark picture:~\cite{U49Bach29D33ALL2019.01, U54D40ccbb2020.03, D34ccbb2020.05, GroundALL0.2020.10, U56D38ALL.2021.01, Excccbb1.2021.24, GroundALL4.2022.07} (all  predictions);~\cite{U48D37ccbb2019.06} (ground states only);~\cite{U55D41ccbb2020.04} (2S);~\cite{Groundccbb4.2021.07} (2S, 3S).

\item in other models:~\cite{U54D40ccbb2020.03, Groundccbb1.2020.08, Groundccbb2.2020.09, Groundccbb3.2021.06} (all  predictions);~\cite{ExcGroundccbb5.2022.04} (ground states only);~\cite{ExcGroundccbb1.2020.11} (3S).
\end{itemize}
\end{itemize}

\par
A number of other conclusions can be drawn from this data:
\begin{itemize}

\item Predictions of Refs.~\cite{GroundALL2.2021.11, GroundALL3.2021.12} are in full agreement with our results for the $cb\bar c\bar b$ tetraquark;

\item Predictions of Refs.~\cite{U49Bach29D33ALL2019.01, GroundALL4.2022.07} give good agreement for the $cb\bar c\bar b$ and $bb\bar b\bar b$ tetraquarks, but do not agree at all for the $cc\bar c\bar c$ tetraquark;

\item Predictions of Refs.~\cite{U48D37ccbb2019.06, U54D40ccbb2020.03, D34ccbb2020.05, Groundccbb4.2021.07, Excccbb1.2021.24} give partial agreement for the $cc\bar c\bar c$ and $bb\bar b\bar b$ tetraquarks in the diquark-antidiquark picture;

\item Predictions of Refs.~\cite{U54D40ccbb2020.03, Groundccbb1.2020.08, Groundccbb3.2021.06, ExcGroundccbb1.2020.11, ExcGroundccbb5.2022.04} give partial agreement for the $cc\bar c\bar c$ and $bb\bar b\bar b$ tetraquarks in models other than the diquark-antidiquark.
\end{itemize}

\noindent
In addition, throughout comparison of our results with those of other scientific groups shows that:

\begin{itemize}
\item For the $cc\bar c\bar c$ tetraquark our results are generally median: there are many results giving both heavier and lighter masses;

\item For the $cb\bar c\bar b$ tetraquark masses our results  exceed those of other scientific groups for all diquark spins and excitations;

\item For the $bb\bar b\bar b$ tetraquark masses our results are slightly higher than those of most other scientific groups.
\end{itemize}

\noindent
The generally higher values of tetraquark masses predicted in our model originate primarily from taking into account the finite size of the diquark. It results in weakening of the one-gluon exchange potential and, thus, increasing the tetraquark mass.
\par
Note that the authors of Ref.~\cite{U46Bach26D29ALL2018.02} came to an unexpected conclusions. They argue that the ground state of the asymmetric tetraquark ${bb\bar c\bar b}$ may be  stable (its ground states have been studied by us in Refs.~\cite{Savch2020,Savch2021} and were found to be significantly above the fall-apart decay thresholds), and they also expect the $cb\bar c\bar b$ tetraquark to be a narrow state in contradiction with our conclusions.

\section{Conclusions}
\label{SecCon}

\par
Within the framework of the relativistic quark model, we calculated masses of the ground states, radial (up to 3S) and orbital (up to 1D) excitations of the fully-charmed $cc\bar c\bar c$, doubly charmed-bottom $cb\bar c\bar b$ and fully-bottom $bb\bar b\bar b$ tetraquarks. An important feature of our calculations is the consistent account of the relativistic effects and the finite size of the diquark (as it is shown in the Sec.~\ref{SecTheor}), which leads to the weakening of the one-gluon exchange potential due to the form factors of the diquark–gluon interaction. 
\par
A detailed analysis of the calculated mass spectra was carried out. We compared calculated tetraquark masses with the thresholds of the strong fall-apart decays into the meson pairs. As it is shown in Sec.~\ref{SecThr}, most of the tetraquark states lie significantly above the meson pair decay threshold. However, tetraquark states with the smallest widths and, as a result, with the most probability to be observed as narrow states have been identified. An argument is given why the excited states in general can be narrow, despite the large phase space.
\par
It should be noted that the mass of the narrow state X(6900) recently discovered in the di-$J/\psi$ pair production (LHCb 2020~\cite{U10-ccLHCb2020}, CMS 2022~\cite{ccCMS2022}, ATLAS 2022~\cite{ccATLAS2022}) agrees  well with our prediction for the masses of the fully-charmed tetraquark excited (2S, 1D) states. According to the calculations, several candidates for the interpretation of this state are proposed. Candidates are also identified for all other recently discovered states such as X(6200) (ATLAS), X(6400) (LHCb), X(6600) (CMS, ATLAS), X(7200) (LHCb, ATLAS), X(7300) (CMS).
\par
In conclusion, we note that experimental searches for fully-heavy tetraquarks are currently ongoing and should be continued. Therefore, it can be expected that new experimental candidates will appear in the near future.

\vspace{6pt} 

\authorcontributions{Authors contributed equally to the preparation of the manuscript. All authors have read and agreed to the published version of the manuscript.}

\funding{The work of Elena M. Savchenko was funded in part by the Foundation for the Advancement of Theoretical Physics and Mathematics ``BASIS'' grant number 22-2-10-3-1.}

\acknowledgments{The authors are grateful to D. Ebert for very fruitful and pleasant collaboration in developing the diquark-antidiquark model of tetraquarks. We are grateful to A.V. Berezhnoy for useful discussions.}

\conflictsofinterest{The authors declare no conflict of interest.}

\begin{adjustwidth}{-\extralength}{0cm}

\reftitle{References}


\bibliography{Symmetry2022}

\end{adjustwidth}
\end{document}